\documentclass[12pt,a4paper]{article}

\usepackage[utf8]{inputenc}
\usepackage{amsmath}
\usepackage{amsfonts}
\usepackage{amssymb}
\usepackage{bbold}
\usepackage[dvips,letterpaper,margin=1.00in,bottom=1.00in]{geometry}
\usepackage{mathrsfs}
\usepackage{caption}
\usepackage{subcaption}
\usepackage[T1]{fontenc}
\usepackage[utf8]{inputenc}
\usepackage{authblk}
\usepackage{scrextend}
\usepackage{footmisc}
\usepackage{indentfirst}
\usepackage{physics}
\usepackage[english]{babel}
\usepackage{pdflscape}
\usepackage[noadjust]{cite}
\usepackage[toc]{appendix}
\usepackage{setspace}
 
\usepackage{graphicx}

\title{Sections and Chapters}
\usepackage{cite}
\title{Analytic Solutions of Scalar Field Cosmology, Mathematical Structures for Early Inflation and Late Time Accelerated Expansion }
\author{Medine Ildes$^{1,}$\footnote{e-mail:medine.ildes@boun.edu.tr}}
\author{Metin Arik$^{1,}$\footnote{e-mail: metin.arik@boun.edu.tr}}
\affil{$^{1}$Department of Physics, Bogazici University, Bebek, Istanbul, Turkiye}

\begin{document}

\maketitle

\begin{abstract}
\indent We study the most general cosmological model with real scalar field which is minimally coupled to gravity. Our calculations are based on Friedmann-Lemaitre-Robertson-Walker (FLRW) background metric. Field equations consist of three differential equations. We switch independent variable from time to scale factor by change of variable $\dot{a}/a=H(a)$. Thus a new set of differential equations are analytically solvable with known methods. We formulate Hubble function, the scalar field, potential and energy density when one of them is given in the most general form. $a(t)$ can be explicitly found as long as methods of integration techniques are available.  We investigate the dynamics of the universe at early times as well as at late times in light of these formulas. We find mathematical machinery which turns on and turns off early accelerated expansion. On the other hand late time accelerated expansion is explained by cosmic domain walls. We have compared our results with recent observations of type Ia supernovae by considering the Hubble tension and absolute magnitude tension. Eighty-nine percent of present universe may consist of domain walls while rest is matter.
\end{abstract}
\section{Introduction}
\indent The scalar field plays an important role in many parts of modern physics. Its usage in cosmology was seen firstly in Nordstrom's studies after Newton's gravity which has a scalar potential field. Although he introduced scalar theory of gravity \cite{nordstrom1912relativitatsprinzip,nordstrom1913theorie,nordstrom1913trage,nordstrom1914uber} in 1912-1913, none of them have been verified by observation \cite{willenbrock1982cosmology}. Then in 1916 Einstein's  theory of general gravity was established. This is a purely tensor theory. Seeds of some alternative theories which in cooperates a scalar field were conceived by Jordan \cite{jordan1955schwerkraft} and Dirac \cite{dirac1938new}.   \\  \\
\indent The standard model of cosmology has the flatness problem, the horizon problem and the monopole problem. In the beginning of 1980's A.H.Guth, A.D.Linde had established inflationary cosmology to solve these problems. In these studies it has been shown that one or more scalar fields drive the early phase of accelerated expansion.  \cite{guth1981inflationary,linde1982new}\\ \\
\indent In 1989-1990 a wide variety of different astrophysical observations have shown that the expansion of the universe is accelerating. \cite{riess1998observational,perlmutter1999measurements,tonry2003cosmological}. According to standard cosmology this behaviour is explained by contribution of dark energy ($\sim 68\%$), dark matter ($\sim 27\%$) and baryonic matter ($\sim 5\%$) to the total density parameter. A Simple explanation of dark energy just as a cosmological constant in standard model of cosmology is problematic \cite{weinberg1989cosmological}. Hence to explain dark energy many different studies have been developed by using scalar fields similar to early inflationary theories. All these models are widely explained in the review article \cite{copeland2006dynamics}. The second main part of the universe consists of dark matter. This still keeps its secrets. It has not been explained  properly yet. A scalar field is again a candidate to reveal its nature \cite{magana2012brief}. \\ \\
\indent Domain walls differentiate among various candidates for dark energy. They supply the required accelerated expansion with negative pressure $p=-(2/3)\rho$. Altough cosmic fluids with a negative equation of state have an imaginary sound speed there have been several studies indicating that cosmic walls are not ruled out in cosmology \cite{battye1999domain,conversi2004domain,friedland2003domain,del2004late,
avelino2011domain,kirillov2020domain}. To include them in cosmology is very appealing because they appear in a field theory which has spontaneously broken discrete symmetries \cite{vachaspati2006kinks}.  \\ \\
\indent The field equations which govern the universe are ordinary differential equations. To be able to solve them many different approcahes have been developed. One of them is the dynamical systems methods in which stabiliy analysis of systems of nonlinear differential equations are investigated. Detailed studies have been performed by this method in \cite{wainwright1997dynamical,coley2003dynamical,bohmer2017dynamical,bahamonde2018dynamical} . Other methods are based on assumptions or approximations. The "slow-roll approximation" is the most common one which is applied in scenarios of the inflationary universe \cite{steinhardt1984prescription,liddle1992cobe}. Lastly, the generating function method is proposed as a method which gives exact solution of the field equations in \cite{kruger2000another}. Some of searches for exact solutions of the field equations where the scalar field is minimally cooupled to gravity have been presented in\cite{fomin2017exact,beesham2015exact,makarenko2012exact,pintus2018mathematical}. \\ 
\\ \indent In this article we have three main purposes. The first is solving field equations analytically. The second is finding a mathematical machinery which causes to turn on and to turn of accelerated expansion in early universe. Last is explaining late time accelerated expansion without dark energy. In section 2 we introduce a mathematical tool which is a change of variable. Thus the field equations are converted to a new set of differential equations. In section 3 we exactly solve this new set of equations and presented solutions in four different forms. In section 4 we investigate single-component universes. In section 5 we examine two-component universes and we find a new exotic matter which causes mathematical mechanisms which turns on and turns off accelerated expansion in an early universe. In section 6 we show that a universe which contains matter and cosmic walls results in accelerated expansion. Furthermore we have compared our results with supernova Ia data. Results are quite satisfactory. Then we examined dark energy dominated universe with the same procedure. Our discussion is given in the conclusion.
\section{Field equations}
\subsection{Original form}
\indent Action of general relativity with scalar field and the cosmological constant is given by
\begin{align}
S=\int \sqrt{-g}[\dfrac{1}{2\kappa}(R-2\Lambda)-\frac{1}{2}g^{\mu \nu}\partial _{\mu} \phi \partial _{\nu}\phi -V(\phi)+ \mathcal{L}_{M}]
\end{align}
\\where $R$ is the Ricci scalar and $\kappa=8\pi Gc^{-4}$. We will use FLWR metric with space dominant metric sign $(-,+,+,+)$ and units with $\hbar =1$, $c=1$
\begin{align}
ds^{2}=-dt^{2}+a^{2}(t)[\frac{dr^{2}}{1-k\dfrac{r^{2}}{L^{2}}}+r^{2}(d\theta ^{2}+\sin^{2}\theta d\varphi ^{2})].
\end{align}
\\where $k=-1,0,1$ and $a(t)=\dfrac{R(t)}{R(t_{0})}$ is normalized scale factor, with the convention $a(t_{0})=1$, $L=R(t_{0})$ and $r$ has dimension of lenght.
\\Energy-momentum tensor for the field is defined as
\begin{align}
T_{\mu\nu}=\dfrac{-2}{\sqrt{-g}}\dfrac{\delta S_{\phi}}{\delta g^{\mu\nu}} 
\end{align} 
\\where $T^{\mu\nu}=\{\rho, p,p,p\}$. \\
 \indent In standard cosmology field equations have been found as
\begin{align}
\dfrac{\dot{a}^{2}}{a^{2}}=\dfrac{8\pi G}{3}\rho+\dfrac{\Lambda}{3}-\dfrac{k}{L^{2}a^{2}} ,\\
\dfrac{\ddot{a}}{a}=-\dfrac{4\pi G}{3}(\rho+3p)+\dfrac{\Lambda}{3} 
\end{align}
\\where $\rho =\rho _{ord}+\rho _{\phi}$ and $p =p _{ord}+p _{\phi}$ are known as  Einstein equations. $ord$ stands for ordinary and represents matter-energy distribution placed in Einstein equations by hand as a function the scale size of the universe.\\
\\In addition, variation of the Lagrangian density with respect to the field $\phi$ gives another equation
\begin{align}
\ddot{\phi}+3H\dot{\phi}+\dfrac{dV}{d\phi}=0.
\end{align}
\\ For scalar field dominated universe we have
\begin{align}
\rho=\dfrac{\dot{\phi}^{2}}{2}+V(\phi) ,\\
p=\dfrac{\dot{\phi}^{2}}{2}-V(\phi)
\end{align}
\\where we assume that $\phi$ is a function only of $t$. Then equation (6) is equivalent to continuity equation which is given by
\begin{align}
\dot{\rho}+3\dfrac{\dot{a}}{a}(\rho+p)=0.
\end{align}
\indent We name our potential as an effective potential in the sense that it may reflect another more basic physical theory. We will call equation (4) as the first Einstein equation, equation (5) as the second Einstein equation and the equation (6) as the $\phi $ equation.
\subsection{New form of field equations}
\indent If independent variable does not appear explicitly in the differential equation one can define a new variable in terms of dependent variables so that the order of the differential equation can be reduced by one \cite{rainville1989elementary}. In our field equations independent variable is $"t"$. We define our new dependent variable as Hubble function
\begin{align}
H(a)=\dfrac{\dot{a}}{a} .
\end{align}
Thus our new independent variable becomes the scale factor \emph{a}. For this reason we write all other variables in terms of the new variable.
\begin{align}
\phi=\phi(a) \hspace{20pt} \text{and} \hspace{20pt} V(\phi)=V(a) .
\end{align}
\\Expressions for derivatives of $a$ and $\phi$ with respect to time in terms of derivatives with respect to new independent variable $a$ are given in appendix A. In addition $\Lambda$ can be included in $V(\phi)$ so we will not carry it anymore.
\\\indent One can easily write the field equations, energy density of the scalar field and the pressure of the scalar field as
\begin{gather}
H^{^2}=\dfrac{8\pi G}{3}\rho-\dfrac{k}{L^{2}a^{2}}, \\ \nonumber\\
H^{'}Ha+H^{2}=-\dfrac{4\pi G}{3}(\rho+3p) , \\ \nonumber\\
\phi^{''}a^{2}H^{2}+4\phi^{'}aH^{2}+\phi^{'}a^{2}HH^{'}+V^{'}\dfrac{1}{\phi^{'}}=0 \\ \nonumber\\
\rho (a)=\dfrac{1}{2}(\phi^{'}aH)^{2}+V(a) , \\ \nonumber \\
p(a)=\dfrac{1}{2}(\phi^{'}aH)^{2}-V(a) 
\end{gather}
\\where prime denotes $\dfrac{d}{da}$.\\
\indent When this set of differential equations is solved exactly we will obtain all unknown functions; $H(a)$, $\phi(a)$, $V(a)$, $\rho (a)$, $p(a)$ and the deceleration parameter $q(a)$ as a function of scale factor, \emph{a}. Thus it will be possible to track the dynamical history of the universe backward and forward in time. Indeed in some cases it will be possible to formulate some of these functions as a function of time.
\section{Solution for field equations}
When solving this differential equation set one should be careful. By taking time derivative of first Einstein equation and using the continuity equation we reach the second Einstein equation. Thus by taking time derivative of first Einstein equation and using the second Einstein equation we reach the continuity equation. In addition by substitution $\rho(t)$ and and $p(t)$ in the continuity equation one can reach $\phi$ equation. One of the field equations can be derivable from other two of them. One can combine these 3 equations in 3 different pairs such that when their solutions are plugged in the remaining differential equation it will be satisfied automatically.\\
\indent First combination is the easiest one. We take the first Einstein equation and the $\phi$ equation.
Then we multiply the $\phi$ equation by $\phi^{'}$ and obtain\\
\begin{align}
a^{2}\phi^{'}\phi^{''}H^{2}+4a\phi^{'2}H^{2}+a^{2}\phi^{'2}HH^{'}+V^{'}(a)=0 .
\end{align}
\\ We define
\begin{align}
\gamma(a)=\frac{\phi^{'2}H^{2}a^{2}}{2}
\end{align}
\\to be able to solve the last differential equation. Therefore this equation is converted to
\begin{align}
\gamma^{'}+\dfrac{6}{a}\gamma=-V^{'} .
\end{align}
\\This is a first order linear differential equation and it's solution can be found easily as
\begin{align}
\gamma(a)=\dfrac{1}{a^{6}}[\int_{a_{in}}^{a} (-a^{'6}V^{'}(a^{'}))da^{'}+a^{6}_{in}\gamma(a_{in})] .
\end{align}
\\Hence by rewriting equation (18) we obtain\\
\begin{gather}
\dfrac{\phi^{'2}H^{2}a^{2}}{2}=\gamma (a) ,\nonumber \\
\phi^{'2}H^{2}a^{2}=\dfrac{2}{a^{6}}[\int_{a_{in}}^{a} (-a^{'6}V^{'}(a^{'}))da^{'}+a^{6}_{in}\gamma(a_{in})] .
\end{gather}
\\It is apparent that to be able to solve this field equation one needs the knowledge of one of the following functions; $V(a)$, $H(a)$, $\phi (a)$. There is one more function which can be used as a starting point of calculations. This is the energy density. The relation between $\rho (a)$ and equation (21) will be studied in section 3.4.  \\
\indent We have gone further by plugging energy density into the first Einstein equation
\begin{align}
H^{2}=\frac{8\pi G}{3}[\dfrac{1}{2}\phi^{'2}H^{2}a^{2}+V(a)]-\dfrac{k}{L^{2}a^{2}} ,
\end{align}
\begin{align}
H^{2}=\dfrac{\dfrac{8\pi G}{3}V(a)-\dfrac{k}{L^{2}a^{2}}}{1-\dfrac{4\pi G}{3}\phi^{'2}a^{2}} .
\end{align}
\\We will refer the last equation as our Friedmann equation.
\subsection{Solution for given $V(a)$}
\indent In this section we start our calculations by using our Friedmann equation. Substituting (23) in (21) we obtain
\begin{align}
\phi^{'2}[\dfrac{\dfrac{8\pi G}{3}V(a)-\dfrac{k}{L^{2}a^{2}}}{1-\dfrac{4\pi G}{3}\phi^{'2}a^{2}}]a^{2}=2\gamma (a) .
\end{align}
\\Then one can reach the following results
\begin{gather}
\phi^{'2}=\dfrac{2\gamma (a)}{a^{2}[\dfrac{8\pi G}{3}(V(a)+\gamma (a))-\dfrac{k}{L^{2}a^{2}}]} , \\ \nonumber \\
\phi(a)=\pm\int_{a_{in}}^{a} \sqrt{\dfrac{2\gamma (a^{'})}{a^{'2}[\dfrac{8\pi G}{3}(V(a^{'})+\gamma (a^{'}))-\dfrac{k}{L^{2}a^{'2}}}}da^{'}+\phi_{a_{in}} .
\end{gather}
\\One should decide to pick one of the $\pm$ sign in front of the right side of $\phi(a)$ such that the value of the scalar field increase or decrease as the universe expands. $H(a)$ has been found by using the formula of the scalar field in our Friedmann equation.
\begin{align}
H(a)=\sqrt{\dfrac{8\pi G}{3}(V(a)+\gamma (a))-\dfrac{k}{L^{2}a^{2}}}
\end{align}
\\where
\begin{align}
\gamma(a)=\dfrac{1}{a^{6}}[\int_{a_{in}}^{a} (-a^{'6}V^{'}(a^{'}))da^{'}+a^{6}_{in}\gamma(a_{in})] .
\end{align}
\indent It is apparent that knowledge of the potential energy function $V(a)$ is sufficient to formulate the scalar field $\phi(a)$ and the Hubble function $H(a)$ as an exact solution of the field equations.\\
\indent We would like to mention that the results of this subsection are similar to results of \cite{chimento1996scalar}. They have reduced the differential equations to quadrature problems by writing $V(a)$ in a complicated way. Then exponential potentials and hyperbolic potentials were focused in their examples.\\ 
\subsection{Solution for given $\phi (a)$ }
\indent In some cases one may need to solve the field equations for a specific scalar field. In this calculation $V(a)$ becomes the unknown dependent variable in (24). To be able to go further first we write the solution of the $\phi$ equation as
\begin{align}
\gamma(a)=-V(a)+\dfrac{1}{a^{6}}[6\int_{a_{in}}^{a} V(a^{'})a^{'5}da^{'}+a^{6}_{in}\tilde{\gamma}(a_{in})], \hspace{15pt} \tilde{\gamma}_(a_{in})=\gamma(a_{in})+V(a_{in})
\end{align}
\\where we have applied integration by parts to equation (20). Details of this calculation are given in the appendix.
\subsubsection{Singular case}
\indent Firstly we will investigate the special form of the scalar field which causes this singularity in the denominator of right side of the equation. From (23) we have
\begin{align}
1-\dfrac{4\pi G}{3}\phi ^{'2}a^{2}=0.
\end{align}
\\Therefore
\begin{gather}
\int _{\phi_{in}}^{\phi}d\phi ^{'}=\pm \sqrt{\dfrac{3}{4\pi G}}\int_{a_{in}}^{a}\dfrac{da'}{a^{'}}    
\end{gather}
\begin{gather}
\phi=
\begin{cases}
\sqrt{\dfrac{3}{4\pi G}}ln(\dfrac{a}{a_{in}})+\phi_{in} \hspace{15pt}, \\
-\sqrt{\dfrac{3}{4\pi G}}ln(\dfrac{a}{a_{in}})+\phi_{in}.
\end{cases}
\end{gather}
\\Since $a_{in}\leq a$, $"+"$ sign indicates that the scalar field always increases. On the other hand minus sign implies that one will have positive and decreasing scalar field when  $\phi _{in}$ big enough. \\
\indent One easily obtains the potential by plugging the field into (22)
\begin{align}
V(a)=\dfrac{3}{8\pi G}(\dfrac{k}{L^{2}a^{2}})
\end{align}
\\ Then the Hubble function is formulated just by substitution of $\phi (a)$ and $V(a)$ into the solution of the $\phi$ equation which is given by the (21) 
\begin{gather}
H^{2}(a)=\dfrac{2}{a^{6}}\{\int_{a_{in}}^{a}\dfrac{k}{L^{2}}a^{'3}da^{'}+\dfrac{4\pi G}{3}a^{6}_{in}\gamma (a_{in})\}, \\
H(a)=\sqrt{\dfrac{k}{2L^{2}a^{2}}+\dfrac{8\pi G}{3}\dfrac{a^{6}_{in}\tilde{\gamma} (a_{in})}{a^{6}}}, \hspace{20pt} \gamma (a_{in})=\dfrac{3k}{16\pi GL^{2}a^{2}_{in}}+\tilde{\gamma} (a_{in})
\end{gather}
\\ At first sight one can say that the spatially flat universe is static by choosing $\tilde{\gamma}(a_{in})=0$. However this statement is incorrect because it is incomplete. Firstly we would like to remind that our choice at the change of variable $\dfrac{\dot{a}}{a}=H(a)$ works only for the dynamic universes where $H(a)\neq 0$. Secondly by using the first Einstein equation one can easily deduce that the static and spatially flat universe must be empty. Therefore complete and correct interpretation says that the spatially flat and dynamic universes have time varying energy density. \\
\indent Considering the solution for $k=0$, it is seen from the equation (33), $V=0$ for spatially flat universe. Thus we jump back to equation (19) and it turns to
\begin{gather}
\gamma^{'}+\dfrac{6}{a}\gamma=0  \\
\int_{\gamma_{in}}^{\gamma}\dfrac{d\gamma^{'}}{\gamma^{'}}=-6\int_{a_{in}}^{a}\dfrac{da^{'}}{a^{'}} \nonumber \\ \nonumber\\
ln\gamma-ln\gamma_{in}=-6(lna-lna_{in}) \nonumber \\
\gamma(a)=\dfrac{\tilde{\gamma}(a_{in})}{a^{6}} \hspace{15pt} \text{and} \hspace{15pt}\tilde{\gamma}(a_{in})=\gamma (a_{in})a^{6}_{in}.
\end{gather}
\\Hence by plugging $\gamma$ and $\phi$ in (18) we have obtain the Hubble function as
\begin{align}
H=\sqrt{\dfrac{8\pi G}{3}\dfrac{\tilde{\gamma}(a_{in})}{a^{6}}}
\end{align}
\subsubsection{Nonsingular case}
\indent In this case we investigate general form of the scalar field where $\phi (a)\neq \sqrt{\dfrac{3}{4\pi G}}ln(a)$. We have start this case by using (29) in (25)
\begin{align}
\phi^{'2}=\dfrac{2\{-V(a)+\dfrac{1}{a^{6}}[6\int_{a_{in}}^{a} V(a^{'})a^{'5}da^{'}+a^{6}_{in}\tilde{\gamma}(a_{in})]\}}{a^{2}\{\dfrac{8\pi G}{3a^{6}}[6\int_{a_{in}}^{a} V(a^{'})a^{'5}da^{'}+a^{6}_{in}\tilde{\gamma}(a_{in})]-\dfrac{k}{L^{2}a^{2}}\}} .
\end{align}
\\To be able to calculate the potential $V(a)$, one should define a new function
\begin{align}
\alpha (a)=6\int_{a_{in}}^{a} V(a^{'})a^{'5}da^{'}+a^{6}_{in}\tilde{\gamma}(a_{in}) \hspace{25pt} V(a)=\dfrac{\alpha^{'}}{6a^{5}}, 
\end{align}
\\where $\alpha(a_{in})=a^{6}_{in}\tilde{\gamma}_{in}$. Then (39) turns into a first order linear differential equation which is obtained as
\begin{align}
\alpha ^{'}+(8\pi Ga\phi^{'2}-\frac{6}{a})\alpha=(\dfrac{3ka^{5}}{L^{2}})\phi ^{'2} .
\end{align}
\\The solution is found as
\begin{align}
\alpha (a)=exp[\int_{a_{in}}^{a}(\dfrac{6}{a^{'}}-8\pi Ga^{'}\phi^{'2})da^{'}]\Big\{\alpha(a_{in})+\int_{a_{in}}^{a}exp[\int_{a_{in}}^{a^{'}}(-\dfrac{6}{a^{''}}+8\pi Ga^{''}\phi^{'2})da^{''}](\dfrac{3ka^{'5}\phi^{'2}}{L^{2}})da^{'}\Big\} .
\end{align}
\\Then according to relation (40) the potential $V(a)$ is found as
\begin{gather}
V(a)=\dfrac{\alpha^{'}}{6a^{5}} ,  \nonumber \\
V(a)=(1-\dfrac{4\pi G}{3}a^{2}\phi^{'2})e^{\lambda(a)}\beta (a)+\dfrac{k\phi^{'2}}{2L^{2}}, \\
\lambda(a)=-\int_{a_{in}}^{a}8\pi Ga^{'}\phi^{'2}da^{'},\\
\beta (a)=\alpha (a_{in})+\int_{a_{in}}^{a}e^{-\lambda(a^{'})}\dfrac{3ka^{6}_{in}\phi^{'2}}{L^{2}a^{'}}da^{'}].
\end{gather}
\\$H(a)$ has been found by substituting this potential and the specific scalar field into our Friedmann equation
\begin{gather}
H^{2}=\dfrac{\dfrac{8\pi G}{3}V(a)-\dfrac{k}{L^{2}a^{2}}}{1-\dfrac{4\pi G}{3}\phi^{'2}a^{2}} , \nonumber \\
H(a)=\sqrt{\dfrac{8\pi G}{3a_{in}^{6}}e^{\lambda(a)}\beta (a)-\dfrac{k}{L^{2}a^{2}}}
\end{gather}
\\where $\lambda (a)$ and  $\beta (a)$ are given by (44) and (45). Hence for a specific scalar field exact solution of the field equations are given by the last two equations \\
\indent Both these two cases have a common physical result. If there is only scalar field without any kinds of matter except the dark energy this universe has dynamic behaviour as a result of it being curved by the scalar field. \\
 \indent For $k=0$ the solution changes. Equation (41) is solved as
 \begin{gather}
 \alpha^{'}=(\dfrac{6}{a}-8\pi Ga\phi^{'2})\alpha \\
 \int_{\alpha_{in}}^{\alpha}\dfrac{d\alpha^{'}}{\alpha ^{'}}=\int_{a_{in}}^{a}(\dfrac{6}{a^{'}}-8\pi Ga^{'}\phi^{'2}(a^{'}))da^{'} \nonumber \\
 ln\alpha-ln\alpha_{in}=(lna^{6}-lna^{6}_{in})-\int_{a_{in}}^{a}8\pi Ga^{'}\phi^{'2}(a^{'})da^{'} \nonumber \\
 \dfrac{\alpha}{\alpha_{in}}=\dfrac{a^{6}}{a^{6}_{in}}exp[-\int_{a_{in}}^{a}8\pi Ga^{'}\phi^{'2}(a^{'})da^{'}] \nonumber \\
 \alpha=\dfrac{\alpha_{in}a^{6}}{a^{6}_{in}}exp[-\int_{a_{in}}^{a}8\pi Ga^{'}\phi^{'2}(a^{'})da^{'}].
 \end{gather}
\\Then we formulate the potential by using equation (40)
\begin{gather}
V(a)=\dfrac{\alpha^{'}}{6a^{5}} \nonumber \\
V(a)=\dfrac{\alpha_{in}}{a^{6}_{in}}[1-\dfrac{4\pi G}{3}a^{2}\phi^{'2}(a)]e^{\lambda (a)}
\end{gather} 
\\where $\lambda (a)$ is given by equation (44). The Hubble function is found just by substuting the potential and the scalar field into our Friedmann equation
\begin{align}
H(a)=\sqrt{\dfrac{8\pi G}{3}\dfrac{\alpha_{in}}{a_{in}^{6}}e^{\lambda (a)}},
\end{align}
\\where $\lambda (a)$ is given in (44).
\subsection{Solution for given $H(a)$}
\indent In this section we will start our calculations by rewriting our Friedmann equation in the form
\begin{align}
H^{2}=\frac{8\pi G}{3}(\dfrac{1}{2}\phi^{'2}H^{2}a^{2}+V(a))-\dfrac{k}{L^{2}a^{2}} . \nonumber
\end{align}
\\This equation is easily converted to 
\begin{align}
\dfrac{\phi ^{'2}a^{2}H^{2}}{2}=\frac{3}{8\pi G}(H^{2}+\dfrac{k}{L^{2}a^{2}})-V(a).
\end{align} 
\\Therefore one can recognize the first term on the left side of equation (51) as $\gamma(a)$ which is the variable found as a solution of the $\phi$ equation at the beginning of the section 3. Hence (51) turns into the following form
\begin{align}
\gamma (a)=\frac{3}{8\pi G}(H^{2}+\dfrac{k}{L^{2}a^{2}})-V(a) .
\end{align}
\\By using the last form of $\gamma (a)$ which is formulated in (29)
\begin{align}
\gamma(a)=-V(a)+\dfrac{1}{a^{6}}[6\int_{a_{in}}^{a} V(a^{'})a^{'5}da^{'}+a^{6}_{in}\tilde{\gamma}(a_{in})]\nonumber.
\end{align}
\\So we have obtained
\begin{align}
-V(a)+\dfrac{1}{a^{6}}[6\int_{a_{in}}^{a} V(a^{'})a^{'5}da^{'}+a^{6}_{in}\tilde{\gamma}(a_{in})]=[\frac{3}{8\pi G}(H^{2}+\dfrac{k}{L^{2}a^{2}})-V(a)] .
\end{align}
\indent As we have done in the previous section we now find the potential energy. The last equation can be easily solved so that $\alpha (a)$ 
\begin{align}
\alpha (a)=6\int_{a_{in}}^{a} V(a^{'})a^{'5}da^{'}+a^{6}_{in}\tilde{\gamma}(a_{in}) \hspace{20pt} \text{so} \hspace{20pt} V(a)=\dfrac{\alpha^{'}}{6a^{5}} . \nonumber
\end{align}
Therefore $\alpha (a)$ is found algebraically from equation (53) as
\begin{align}
\alpha (a)=\dfrac{3a^{6}}{8\pi G}(H^{2}+\dfrac{k}{L^{2}a^{2}}) .
\end{align}
\\As a result the potential energy is calculated as
\begin{align}
V(a)=\dfrac{3}{8\pi G}[(H^{2}+\dfrac{k}{L^{2}a^{2}})+\dfrac{a}{3}(HH^{'}-\dfrac{k}{L^{2}a^{3}})].
\end{align}
\\The scalar field is found by substituting the potential into equation (51) as
\begin{align}
\phi (a)=\pm\int_{a_{in}}^{a} \sqrt{\dfrac{1}{4\pi GaH^{2}}(-HH^{'}+\dfrac{k}{L^{2}a^{'3}})}  da^{'}+\phi (a_{in}) .
\end{align}
\\Therefore last equations can be used to construct the scalar field and the potential for a given Hubble function.
\subsection{Solution for given $\rho (a)$}
\indent When one starts the calculations with one of the following functions $V(a)$, $\phi(a)$, $H(a)$ one can end up with some unusual forms of the energy density. To avoid this possibility one should start the calculations for desired energy density. It is written in terms of our new independent variable "\emph{a}" as
\begin{align}
\rho (a)=\dfrac{1}{2}(\phi^{'}aH)^{2}+V(a) . \nonumber 
\end{align}
One can recognize the first term on the right side of this equation as $\gamma (a)$. Hence we obtain
\begin{align}
V(a)=\rho (a)-\gamma (a) \hspace{15pt} \text{and} \hspace{15pt} \gamma (a)=\dfrac{(\phi^{'}aH)^{2}}{2}  .
\end{align}
\\Then we substitute this into the $\phi$ equation 
\begin{align}
\gamma^{'}+\dfrac{6}{a}\gamma=-V^{'} \nonumber
\end{align}
\\and we obtain 
\begin{align}
\gamma (a)=- \dfrac{a}{6}\rho {'}.
\end{align}
\\By using the definition of $\gamma (a)$ we  also obtain 
\begin{align}
\phi ^{'2}=\dfrac{2\gamma (a)}{a^{2}H^{2}}.
\end{align}
\\The Hubble function is known just by inserting the energy density into the original form of the first Einstein equation
\begin{align}
H(a)=\sqrt{\dfrac{8\pi G}{3}\rho-\dfrac{k}{L^{2}a^{2}}} .
\end{align}
\\Therefore the scalar field is formulated as
\begin{align}
\phi (a)=\pm \int_{a_{in}}^{a} \sqrt{\dfrac{- \dfrac{a^{'}}{3}\rho {'}}{a^{'2}[\dfrac{8\pi G}{3}\rho-\dfrac{k}{L^{2}a^{'}}]}}da^{'}+\phi (a_{in}) .
\end{align}
\\The potential energy is written by substitution of (58) into (57) as 
\begin{align}
V(a)=\rho (a)+\dfrac{a}{6}\rho {'} .
\end{align}
\\Exact solution of the field equations for a desired energy density are given by the formulas in (60-62).
\section{Single Component Universes}
\indent We present some general solutions for a universe which has a single component. This purpose is easily achieved for a given $\rho(a)$ in section (4.1) and for a given $V(a)$ in section (4.2). In addition we have performed calculations for both a curved universe and a spatially flat universe. Therefore one can see the effect of curvature term in dynamics of the universe.
\subsection{General solution for $\rho(a)=\dfrac{\rho_{n}}{a^{n}}$}
\indent To satisfy the weak energy condition $\rho\geq 0$ and $\rho +p\geq 0$ one should start calculations with a given energy density.
\subsubsection{$k\neq 0$}
\indent We begin this subsection by taking the energy density as in the form of perfect fluid
\begin{align}
\rho=\dfrac{\rho_{n}}{a^{n}}.
\end{align}
\\Then by applying the procedure which is explained in section (3.4) we immediately obtain $H(a)$, $V(a)$, $\phi (a)$, $p(a)$, $q(a)$ as
\begin{gather}
H(a)=\sqrt{\dfrac{8\pi G\rho_{n}}{3a^{n}}-\dfrac{k}{a^{2}}},\\
V(a)=\dfrac{(6-n)}{6}\dfrac{\rho_{n}}{a^{n}},\\
\phi(a)=\pm\sqrt{\dfrac{n}{8\pi G}}\{ln(\dfrac{a}{a_{in}})+\dfrac{1}{1-n/2}ln[\dfrac{1}{b}(1+\sqrt{1-\dfrac{3ka^{n-2}}{8\pi G\rho_{n}}})]\}+\phi(a_{in}),\\
b=(1+\sqrt{1-\dfrac{3ka_{in}^{n-2}}{8\pi G\rho_{n}}}), \nonumber \\
p(a)=(\dfrac{n-3}{3})\dfrac{\rho _{n}}{a^{n}},\\
q(a)=\dfrac{4(n-2)\pi G\rho _{n}}{8\pi G\rho_{n}-3ka^{n	-2}}.
\end{gather}
\\First, we interpret these formulas generally. When we apply the boundaries on equation of state $\nu=\dfrac{p}{\rho}$ we obtian
\begin{align}
-1\leq \nu \leq 1, \hspace{15pt} -1\leq \dfrac{n-3}{3}\leq 1,\hspace{15pt}0\leq n \leq6
\end{align}
\\Therefore all exotic fluids with energy density in the form of $\dfrac{\rho_{n}}{a^{n}}$ have $0\leq n \leq 6$. This condition also makes the potential non-negative. Then special cases pops up immediately for $n=0,2,6$. Furthermore we would like to add one more comment. The second term on the right side of (66) is always real. This is easily recognized when one writes the related components in terms of cosmological density parameters. \\
\indent Case $n=0$ corresponds to constant energy density $\rho=\rho_{0}$. One presents related functions for more comments,
\begin{gather}
H(a)=\sqrt{\dfrac{8\pi G\rho_{0}}{3}-\dfrac{k}{a^{2}}},\\
V(a)=\rho_{0},\\
\phi (a)=0,\\
p(a)=-\rho_{0},\\
q(a)=-1+\dfrac{3k}{3k-8\pi G\rho_{0}a^{2}}.
\end{gather}
\\Furthermore $a(t)$ can be formulated by the following steps
\begin{gather}
\int_{0}^{t}dt^{'}=\int_{a_{in}}^{a}\dfrac{da^{'}}{a^{'}H(a^{'})},\\ \nonumber \\
a(t)=(\dfrac{a_{in}}{2}+\dfrac{\sqrt{\mu^{2}a_{in}^{2}-k}}{2\mu})e^{\mu t}+(\dfrac{a_{in}}{2}-\dfrac{\sqrt{\mu^{2}a_{in}^{2}-k}}{2\mu})e^{-\mu t}, \hspace{15pt}\mu=\sqrt{\dfrac{8\pi G\rho_{0}}{3}}.
\end{gather}
\\Since the scalar field is zero in this case, our solutions reduce to solutions of standard cosmology with dark energy. According to results given in  (70) and (76), to have a dynamic universe with real Hubble function and real scale factor cosmological constant must be big enough to overcome the smallness of the universe;
\begin{align}
\dfrac{8\pi G\rho_{0}}{3} > \dfrac{k}{a_{in}^{2}}.
\end{align}
\\This is the mathematical reason which explains the big value of the cosmological constant in the early universe according to the standard model.\\
\indent Case $n=2$ creates a singularity in the scalar field as seen in (66). Thus we have calculated $\phi (a)$ separately and we have found it as
\begin{align}
\phi(a)=\pm\sqrt{\dfrac{2\rho_{2}}{8\pi G\rho_{2}-3k}}ln(\dfrac{a}{a_{in}})+\phi(a_{in}).
\end{align} 
\\Furthermore $a(t)$ is
\begin{align}
a(t)=\tau t+a_{in},\hspace{20pt}\tau=\sqrt{\dfrac{8\pi G\rho_{2}}{3}-k}
\end{align}
\\which is consistent with (68) which tells us that for $n=2$, the universe expands with constant speed.\\
\indent Case $n=6$ requires special attention. Potential becomes $V=0$ and equation of state becomes $\nu=\dfrac{p}{\rho}=1$. This case corresponds to a massless scalar field. \\
\subsubsection{$k=0$}
\indent When we study the spatially flat universe, nature of the scalar field changes. As a result of this change we can formulate the potential as a function of the scalar field.
\begin{align}
\phi (a)=\pm\sqrt{\dfrac{n}{8\pi G}}ln(\dfrac{a}{a_{in}})+\phi (a_{in}).
\end{align}
\\Therefore one can formulate the scale factor and hence the potential as a function of the scalar field as
\begin{gather}
a=a_{in}exp[\pm\sqrt{\frac{8\pi G}{n}}(\phi-\phi(a_{in}))],\\
V(a)=(\dfrac{6-n}{6})\dfrac{\rho_{n}}{a_{in}^{n}}exp[\mp\sqrt{8\pi nG}(\phi-\phi(a_{in}))].
\end{gather}
\\Furthermore we can find $a(t)$ by
\begin{gather}
\int_{0}^{t}dt^{'}=\int_{a_{in}}^{a}\dfrac{da^{'}}{a^{'}H(a^{'})},\nonumber \\
a(t)=(n\sqrt{\dfrac{2\pi G\rho_{n}}{3}}t+a_{in}^{n/2})^{2/n}
\end{gather}
\\where there is a singularity in the case $n=0$. This case corresponds to the standard model with cosmological constant. Hence 
\begin{gather}
\int_{0}^{t}dt^{'}=\int_{a_{in}}^{a}\dfrac{da^{'}}{a^{'}H(a^{'})},\nonumber \\
a(t)=a_{in}e^{\mu t},\hspace{20pt}  \mu=\sqrt{\dfrac{8\pi G\rho_{0}}{3}}.
\end{gather}
\\As it is seen there is no constraint on this constant energy density. It can be big as well as it can be small.\\
\subsection{General solution for $V(a)=\dfrac{V_{n}}{a^{n}}$}
\indent To start calculations with given $V(a)$ is more fundamental. After getting intuition about the form of the potential which is required for the perfect fluid, we continue our work by choosing the potential. Results are important because they are surprisingly different than section (4.1).
\subsubsection{$k\neq 0$}
\indent We have plugged in $V(a)=\dfrac{V_{n}}{a^{n}}$ in the formulas given by (26-28) and we have obtained
\begin{gather}
\gamma(a)=\dfrac{1}{a^{6}}[\dfrac{nV_{n}}{6-n}a^{6-n}+a^{6}_{in}\tilde{\gamma}(a_{in})], \hspace{15pt} \tilde{\gamma}(a_{in})=-\dfrac{nV_{n}}{6-n}a_{in}^{-n}+\gamma(a_{in}),\\
H(a)=\sqrt{\dfrac{16\pi GV_{n}}{(6-n)a^{n}}+\dfrac{8\pi Ga^{6}_{in}\tilde{\gamma} (a_{in})}{3a^{6}}-\dfrac{k}{L^{2}a^{2}}},\\
\phi (a)=\pm\sqrt{6}\int_{a_{in}}^{a}\sqrt{\dfrac{nV_{n}a^{'6}+(6-n)a^{6}_{in}\tilde{\gamma} (a_{in})a^{'n}}{48\pi GV_{n}a^{'8}+(n-6)(3ka^{'4}-8\pi Ga^{6}_{in}\tilde{\gamma} (a_{in}))a^{'n+2}}}da^{'}+\phi (a_{in}),\\
\rho(a)=\dfrac{6V_{n}}{(6-n)a^{n}}+\dfrac{a^{6}_{in}\tilde{\gamma} (a_{in})}{a^{6}},\\
p(a)=\dfrac{2(n-3)V_{n}}{(6-n)a^{n}}+\dfrac{a^{6}_{in}\tilde{\gamma} (a_{in})}{a^{6}},\\
q(a)=\dfrac{8\pi G[2(6-n)a^{6}_{in}\tilde{\gamma} (a_{in})a^{n}+3(n-2)V_{n}a^{6}]}{48\pi GV_{n}a^{6}+(n-6)(3ka^{4}-8\pi Ga^{6}_{in}\tilde{\gamma} (a_{in}))a^{n}}.
\end{gather}
\\For all $n$ there is the term proportional to $a^{-6}$ in energy density. Therefore not only for zero potential but also for each potential, universe contains the stiff fluid.\\
\indent For case $n=0$ one should perform the calculations by starting from equation (19),
\begin{gather}
\int_{\gamma (a_{in})}^{\gamma}\dfrac{d\gamma}{\gamma}=-\int_{a_{in}}^{a}\dfrac{6}{a^{'}}da^{'},\\
\gamma (a)=\dfrac{\gamma(a_{in})a_{in}^{6}}{a^{6}},
\end{gather}
\begin{gather}
H(a)=\sqrt{\dfrac{8\pi G}{3}(\dfrac{\gamma(a_{in})a_{in}^{6}}{a^{6}}+V_{0})-\dfrac{k}{L^{2}a^{2}}},\\
\phi (a)=\pm\sqrt{6}\int_{a_{in}}^{a}\sqrt{\dfrac{\gamma(a_{in})a_{in}^{6}}{8\pi G(\gamma(a_{in})a_{in}^{6}+V_{0}a^{'6})a^{'2}-3ka^{'6}}}da^{'}+\phi (a_{in}),\\
\rho(a)=\dfrac{\gamma(a_{in})a_{in}^{6}}{a^{6}}+V_{0},\\
p(a)=\dfrac{\gamma(a_{in})a_{in}^{6}}{a^{6}}-V_{0},\\
q(a)=\dfrac{8\pi G[2\gamma(a_{in})a_{in}^{6}-V_{0}a^{6}]}{8\pi G[\gamma(a_{in})a_{in}^{6}+V_{0}a^{6}]-3ka^{4}}.
\end{gather}
\\In contrast to constant energy density case, constant potential differs from cosmological constant case. \\
\indent As it is seen from formulas there is a singularity for $n=6$. Thus we have investigated this case separately:
\begin{gather}
\gamma (a)=\dfrac{1}{a^{6}}[6V_{6}ln(\dfrac{a}{a_{in}})+a^{6}_{in}\gamma(a_{in})],\\
H(a)=\sqrt{\dfrac{8\pi G}{3}[\dfrac{V_{6}+a^{6}_{in}\gamma(a_{in})}{a^{6}}+\dfrac{6V_{6}ln(\dfrac{a}{a_{in}})}{a^{6}}]-\dfrac{k}{L^{2}a^{2}}},\\
\phi(a)=\pm\sqrt{6}\int_{a_{in}}^{a}\sqrt{\dfrac{a^{6}_{in}\gamma(a_{in})+6V_{6}ln(\dfrac{a^{'}}{a_{in}})}{[48\pi GV_{6}ln(\dfrac{a^{'}}{a_{in}})+8\pi G(V_{6}+a^{6}_{in}\gamma(a_{in}))-3ka^{'4}]a^{'2}}}da^{'}+\phi(a_{in}),\\
\rho (a)=\dfrac{a^{6}_{in}\gamma(a_{in})+V_{6}}{a^{6}}+\dfrac{6V_{6}ln(\dfrac{a}{a_{in}})}{a^{6}},\\
p(a)=\dfrac{a^{6}_{in}\gamma(a_{in})-V_{6}}{a^{6}}+\dfrac{6V_{6}ln(\dfrac{a}{a_{in}})}{a^{6}},\\
q(a)=\dfrac{8\pi G[2a^{6}_{in}\gamma(a_{in})-V_{6}+12V_{6}ln(\dfrac{a}{a_{in}})]}{48\pi GV_{6}ln(\dfrac{a}{a_{in}})+8\pi G(a^{6}_{in}\gamma(a_{in})+V_{6})-3ka^{4}}.
\end{gather}
\\Energy density and pressure should be written in the following form\\
\begin{gather}
\rho(a)=\rho_{1}(a)+\rho_{2}(a), \\
\rho_{1}(a)=\dfrac{a^{6}_{in}\gamma(a_{in})}{a^{6}}\hspace{20pt} \rho_{2}(a)=\dfrac{V_{6}}{a^{6}}ln[e(\dfrac{a}{a_{in}})^{6}],\\
p(a)=p_{1}(a)+p_{2}(a),\\
p_{1}(a)=\dfrac{a^{6}_{in}\gamma(a_{in})}{a^{6}}\hspace{20pt} p_{2}(a)=\dfrac{V_{6}}{a^{6}}ln[\dfrac{1}{e}(\dfrac{a}{a_{in}})^{6}].
\end{gather}
\\Thus each component satisfies the continuity equation which is given by (9) according to perfect fluid theorem.\\
\\Furthermore investigation of equation of state for the second part of the fluid is important. First we write the pressure in the following form:\\
\begin{gather}
p_{2}(a)=\dfrac{V_{6}}{a^{6}}ln[\dfrac{1}{e}(\dfrac{a_{in}}{a})^{6}(\dfrac{a}{a_{in}})^{12}],\\
p_{2}(a)=\dfrac{V_{6}}{a^{6}}\{ln[\dfrac{1}{e}(\dfrac{a_{in}}{a})^{6}]+ln(\dfrac{a}{a_{in}})^{12}\}.
\end{gather} 
\\Then equation of state turns into
\begin{align}
\nu_{2}&=\dfrac{p_{2}}{\rho_{2}}, \\
\nu_{2}&=\dfrac{ln[\dfrac{1}{e}(\dfrac{a_{in}}{a})^{6}]+ln(\dfrac{a}{a_{in}})^{12}}{ln[e(\dfrac{a}{a_{in}})^{6}]},\\
\nu_{2}&=-1+\dfrac{12ln(\dfrac{a}{a_{in}})}{1+6ln(\dfrac{a}{a_{in}})}.
\end{align}
\\In addition
\begin{align}
\lim_{a\to a_{in}} \nu_{2}=-1, \hspace{20pt} \lim_{a\to\infty} \nu_{2}=1.
\end{align}
\\This phenomenon says that at the beginning of the universe there was a negative pressure. This pressure was huge because it is proportional to $\dfrac{1}{a^{6}}$. As the universe expands this pressure and the related energy density becomes negligible since both of them proportional to $\dfrac{1}{a^{6}}$. \\
\subsubsection{k=0}
\indent First simplification occurs in the relation between cosmological time and the scale factor of the universe as
\begin{gather}
t=\int_{a_{in}}^{a}\dfrac{da^{'}}{a^{'}H(a^{'})}, \nonumber \\
t=\int_{a_{in}}^{a}\dfrac{da^{'}}{a^{'}\sqrt{\dfrac{8\pi G}{3}(\dfrac{a_{in}^{6}\tilde{\gamma}(a_{in})}{a^{'6}}+\dfrac{6v_{n}}{(6-n)a^{'n}})}}, \\
t=\sqrt{\dfrac{1}{24\pi Ga_{in}^{6}\tilde{\gamma}(a_{in})}}\Big[a^{'3}{}_{2}F_{1}(\dfrac{1}{2},\dfrac{3}{6-n};\dfrac{3}{6-n}+1;-\dfrac{6v_{n}}{(6-n)a_{in}^{6}\tilde{\gamma}(a_{in})}a^{'(6-n)})\Big]_{a_{in}}^{a}.
\end{gather}
\\${}_{2}F_{1}$ is the hypergeometric function 
\begin{align}
{}_{2}F_{1}(1/2,b;b+1;u)=\sum_{n=0}^{\infty}\dfrac{(1/2)_{n}(b)_{n}}{(b+1)_{n}}\dfrac{u^{n}}{n!},
\end{align}
\\where $(b)_{n}$ is the Pochhammer symbol which is defined as
\begin{align}
(b)_{n}=
\begin{cases}
1 \hspace{20pt} \text{if} \hspace{20pt} n=0, \\
b(b+1)(b+2)...(b+n-1),\hspace{20pt} \text{if} \hspace{20pt} n>0. 
\end{cases}
\end{align}
\\ Our condition, $n<6$ for positivity of energy avoids the singularity of the hypergeometric function given in (115).\\
\indent One can go further by choosing initial condition $a_{in}^{6}\tilde{\gamma}(a_{in})=0$. Results are very similar to what we have obtained in section 4.1.2. The scale factor is found as
\begin{align}
a(t)=(2n\sqrt{\dfrac{\pi Gv_{n}}{(6-n)}}t+a_{in}^{n/2})^{2/n}.
\end{align}
\\Positivity condition $n<6$ for energy density makes $a(t)$ real. The scalar field is found as
\begin{align}
\phi (a)=\pm\sqrt{\dfrac{n}{8\pi G}}ln(\dfrac{a}{a_{in}})+\phi(a_{in}).
\end{align}
\\Then
\begin{align}
a=a_{in}exp(\pm \mu \psi), \hspace{15pt} \psi=\phi(a)-\phi(a_{in})\hspace{15pt} \text{and} \hspace{15pt} \mu=2\sqrt{\dfrac{2\pi G}{n}}.
\end{align}
\\Therefore
\begin{align}
V(\psi)=V_{n}a_{in}^{-n}exp(\mp n\mu \psi)\hspace{15pt} \text{or} \hspace{5pt} V(\phi)=V_{n}a_{in}^{-n}exp\{\mp n\mu [\phi(a)-\phi(a_{in})]\}.
\end{align}
\indent In constant potential case where $n=0$ one can obtain explicit form of the scalar field as
\begin{gather}
\phi(a)=\mp\dfrac{1}{2}\sqrt{\dfrac{1}{3\pi G}}ln[\dfrac{1}{ba^{3}}(1+\sqrt{1+\dfrac{V_{0}a^{6}}{\gamma(a_{in})a_{in}^{6}}})]+\phi(a_{in}),\\
b=\dfrac{1}{a_{in}^{3}}(1+\sqrt{1+\dfrac{V_{0}}{\gamma(a_{in})}}).\nonumber
\end{gather}
\\Formulation of $a(t)$ is also possible
\begin{gather}
t=\int_{a_{in}}^{a}\dfrac{da^{'}}{a^{'}\sqrt{\dfrac{8\pi G}{3}(V_{0}+\dfrac{\gamma(a_{in})a_{in}^{6}}{a^{'6}})}}, \\
a(t)=[\dfrac{b^{2}V_{0}e^{2\mu t}-\gamma(a_{in})a_{in}^{6}e^{-2\mu t}}{2bV_{0}}]^{1/3}, \\
b=a_{in}^{3}+\sqrt{a_{in}^{6}+\dfrac{\gamma(a_{in})a_{in}^{6}}{V_{0}}} \hspace{20pt} \text{and} \hspace{20pt} \mu=\sqrt{6\pi GV_{0}}. \nonumber
\end{gather}
\\This case will be investigated in more detail in section 5.3.\\
\indent The case $n=6$ has two simplification for a spatially flat universe. The scalar field has been found as
\begin{gather}
\phi(a)=\pm \dfrac{1}{12}\sqrt{\dfrac{3}{\pi G}}\dfrac{1}{V_{6}}I(a)+\phi(a_{in}), \\
I(a^{'})=x(a^{'})\sqrt{1-\dfrac{V_{6}}{x(a^{'})}}+\dfrac{V_{6}}{2}ln[-\dfrac{2x(a^{'})}{V_{6}}+1+\dfrac{2x(a^{'})}{V_{6}}\sqrt{1-\dfrac{V_{6}}{x(a^{'})}}],\\
x(a^{'})=a_{in}^{6}\gamma(a_{in})+V_{6}[1+6ln(\dfrac{a^{'}}{a_{in}})].
\end{gather}
\\Expression of $a(t)$ is 
\begin{gather}
a(t)=a_{in}exp\Big \{\dfrac{1}{6}[2\Big (erfi^{-1}(\mu t+b)\Big )^{2}-1-\dfrac{a_{in}^{6}\gamma(a_{in})}{V_{6}}]\Big\},\\
\mu=\sqrt{\dfrac{8\pi G}{3}}\dfrac{1}{a_{in}^{3}}, \hspace{20pt} b=erfi[\dfrac{1}{2}\Big (1+\dfrac{a_{in}^{6}\gamma(a_{in})}{V_{6}}\Big )],\\
erfi(\theta)=\dfrac{2}{\sqrt{\pi}}\int_{0}^{\theta} exp(z^{2})dz,
\end{gather}
\\where $erfi(\theta)$ is the \emph{imaginary error function}. When $\theta$ is real $erfi(\theta)$ is real \cite{korotkov2020integrals}. Therefore it's inverse function $erfi^{-1}(\theta)$ becomes real for real $\theta$.
\section{Early epoch of the universe}
\indent There have been many studies which show that the early universe should expand with exponential expansion to be able to reach its size today. Thus our purpose in this section is to explore the mathematical turn on and turn of mechanism to start and to end up exponential expansion. For this reason we have studied three different combinations for curved and spatially flat universes. 
\subsection{Dark energy}
\indent We will search for the universe with energy density in the following form
\begin{align}
\rho(a)=\dfrac{\rho_{n}}{a^{n}}+\rho_{0}.
\end{align}
\subsubsection{$k\neq 0$}
\indent Firstly we take nonzero curvature. We have found $H(a)$, $V(a)$, $\phi(a)$, $p(a)$ and $q(a)$ as
\begin{gather}
H(a)=\sqrt{\dfrac{8\pi G}{3}(\dfrac{\rho_{n}}{a^{n}}+\rho_{0})-\dfrac{k}{a^{2}}},\\
V(a)=\dfrac{(6-n)}{6}\dfrac{\rho_{n}}{a^{n}}+\rho_{0},\\
\phi (a)=\pm\int_{a_{in}}^{a}\sqrt{\dfrac{n\rho_{n}}{8\pi G(\rho_{n}a^{'2}+\rho_{0}a^{'2+n})-3ka^{'n}}}da^{'}+\phi (a_{in}),\\
p(a)=\dfrac{(n-3)}{3}\dfrac{\rho_{n}}{a^{n}}-\rho_{0},\\
q(a)=\dfrac{4\pi G[-2\rho_{0}a^{n}+(n-2)\rho_{n}]a^{2}}{8\pi G[\rho_{n}a^{2}+\rho_{0}a^{2+n}]-3ka^{n}}.
\end{gather}
\\We continue to investigate the dynamics of the early universe for small a as
\begin{align}
\lim_{a\rightarrow 0}q(a)=\dfrac{4\pi G[(n-2)\rho_{n}]a^{2}}{8\pi G\rho_{n}a^{2}-3ka^{n}},\\
\lim_{a\rightarrow 0}q(a)=
\begin{cases}
-1+\dfrac{n}{2} \hspace{20pt} \text{if} \hspace{15pt} 2< n\leq 6,\\
0 \hspace{45pt} \text{if} \hspace{15pt} 0\leq n\leq 2.
\end{cases}
\end{align}
\\One should check the roots of $q(a)$. Since the denominator of the $q(a)\sim H^{2}(a)$ and $H^{2}(a)>0$ we are only interested in numerator of $q(a)$.
\begin{align}
q(a)=0 \hspace{20pt}\Rightarrow \hspace{20pt} a=[\dfrac{(n-2)\rho_{n}}{2\rho_{0}}]^{1/n}
\end{align} 
\\where there is no real and positive root for $n\leq 2$. Hence for $2<n\leq 6$ the universe starts to expand with deceleration and then expansion turns to acceleration. For $0<n\leq 2$ the universe starts with constant Hubble function and it immediately accelerates. In both cases although there are mathematical turn on mechanism to initiate acceleration there is no mathematical turn of mechanism to end acceleration in the this model.\\
\subsubsection{$k=0$}
\indent For $k=0$, behaviour of the deceleration parameter changes as follows
\begin{align}
\lim_{a\rightarrow 0}q(a)=-1+\dfrac{n}{2} \hspace{20pt} \text{if} \hspace{20pt} 0\leq n\leq 6.
\end{align}
\\Root of $q(a)$ is still given by (139). The difference between spatially flat and curved universe occurs just at the beginning of the universe for $n<2$. The universe starts to expand with acceleration.\\
\indent The scalar field can be simplified as 
\begin{gather}
\phi(a)=\mp\dfrac{1}{\sqrt{2\pi Gn}}ln[\dfrac{a^{-n/2}+\sqrt{a^{-n}+\dfrac{\rho_{0}}{\rho_{n}}}}{a_{in}^{-n/2}+\sqrt{a_{in}^{-n}+\dfrac{\rho_{0}}{\rho_{n}}}}]+\phi(a_{in}).
\end{gather}
\\Then one can derive $V(\phi)$ as follows
\begin{gather}
\dfrac{a^{-n/2}+\sqrt{a^{-n}+\dfrac{\rho_{0}}{\rho_{n}}}}{a_{in}^{-n/2}+\sqrt{a_{in}^{-n}+\dfrac{\rho_{0}}{\rho_{n}}}}=exp(-\sqrt{n}\psi), \hspace{15pt} \psi=\pm\sqrt{2\pi G}[\phi(a)-\phi(a_{in})].
\end{gather}
\\Thus for some  specific values of $n$ one obtains
\begin{gather}
V(\psi)=
\begin{cases}
\rho_{0} \hspace{20pt} \text{if} \hspace{20pt} n=6,\\
\dfrac{1}{3}\dfrac{\rho_{4}}{a^{4}(\psi)}+\rho_{0}\hspace{20pt} \text{if} \hspace{20pt} n=4,\\
\dfrac{1}{2}\dfrac{\rho_{3}}{a^{3}(\psi)}+\rho_{0}\hspace{20pt} \text{if} \hspace{20pt} n=3,
\end{cases}\\ \nonumber\\
a(\psi)=
\begin{cases}
\sqrt{2}\{\dfrac{-a_{in}^{2}e^{2\psi}[1-\sqrt{1+a_{in}^{4}\dfrac{\rho_{0}}{\rho_{4}}}+(1+\sqrt{1+a_{in}^{4}\dfrac{\rho_{0}}{\rho_{4}}})e^{4\psi}]}{a_{in}^{4}\dfrac{\rho_{0}}{\rho_{4}}(1+e^{8\psi})-2(2+a_{in}^{4}\dfrac{\rho_{0}}{\rho_{4}})e^{4\psi}}\}^{1/2}\hspace{20pt} \text{if} \hspace{20pt} n=4,\\
2^{2/3}\{\dfrac{a_{in}^{3/2}e^{\sqrt{3}\psi}[-1+\sqrt{1+a_{in}^{3}\dfrac{\rho_{0}}{\rho_{3}}}-(1+\sqrt{1+a_{in}^{3}\dfrac{\rho_{0}}{\rho_{3}}})e^{2\sqrt{3}\psi}]}{a_{in}^{3}\dfrac{\rho_{0}}{\rho_{3}}(1+e^{4\sqrt{3}\psi})-2(2+a_{in}^{3}\dfrac{\rho_{0}}{\rho_{3}})e^{2\sqrt{3}\psi}}\}^{2/3}\hspace{20pt} \text{if} \hspace{20pt} n=3.
\end{cases}
\end{gather}
\indent Formulation of $a(t)$ is possible for $k=0$ as
\begin{gather}
t=\int_{a_{in}}^{a}\dfrac{da^{'}}{a^{'}H(a^{'})},\\
t=\sqrt{\dfrac{3}{8\pi G\rho_{0}}}\dfrac{2}{n}ln[\dfrac{a^{n/2}+\sqrt{a^{n}+\dfrac{\rho_{n}}{\rho_{0}}}}{a^{n/2}_{in}+\sqrt{a^{n}_{in}+\dfrac{\rho_{n}}{\rho_{0}}}}].
\end{gather}
\\For some specific $n$
\begin{gather}
a(t)=[\dfrac{(a_{in}^{3}+\sqrt{a_{in}^{6}+\dfrac{\rho_{n}}{\rho_{0}}})e^{3\mu t}+(a_{in}^{3}-\sqrt{a_{in}^{6}+\dfrac{\rho_{n}}{\rho_{0}}})e^{-3\mu t}}{2}]^{1/3}, \hspace{15pt} \text{if} \hspace{15pt} n=6, \\
a(t)=[\dfrac{(a_{in}^{2}+\sqrt{a_{in}^{4}+\dfrac{\rho_{n}}{\rho_{0}}})e^{2\mu t}+(a_{in}^{2}-\sqrt{a_{in}^{4}+\dfrac{\rho_{n}}{\rho_{0}}})e^{-2\mu t}}{2}]^{1/2}, \hspace{15pt} \text{if} \hspace{15pt} n=4, \\
a(t)=[\dfrac{(a_{in}^{3/2}+\sqrt{a_{in}^{3}+\dfrac{\rho_{n}}{\rho_{0}}})e^{3\mu t/2}+(a_{in}^{3/2}-\sqrt{a_{in}^{3}+\dfrac{\rho_{n}}{\rho_{0}}})e^{-3\mu t/2}}{2}]^{2/3}, \hspace{15pt} \text{if} \hspace{15pt} n=3,
\end{gather}
\\where $\mu=\sqrt{\dfrac{8\pi G\rho_{0}}{3}}$. If one chooses $a_{in}=0$, for $t\ll 1$ $a(t)$ can be approximately written as\\
\begin{align}
a(t)=
\begin{cases}
3^{1/3}(\dfrac{\rho_{6}}{\rho_{0}})^{1/6}(\mu t)^{1/3} \hspace{15pt} \text{if} \hspace{15pt}n=6,\\
2^{1/2}(\dfrac{\rho_{4}}{\rho_{0}})^{1/4}(\mu t)^{1/2} \hspace{15pt} \text{if} \hspace{15pt}n=4, \\
(\dfrac{3}{2})^{2/3}(\dfrac{\rho_{3}}{\rho_{0}})^{1/3}(\mu t)^{2/3} \hspace{15pt} \text{if} \hspace{15pt}n=3.
\end{cases}
\end{align}
\subsection{Cosmic domain walls}
\indent Cosmic domain walls are known with their contribution to energy density with term $\rho\sim 1/a$. Their equation of state is given by $\mu=-2/3$. Dynamics of the universe with two components where one of them is a domain wall are very similar to dynamics of universes with two components where one of them is dark energy. We have taken the energy density in the following form
\begin{align}
\rho(a)=\dfrac{\rho_{w}}{a}+\dfrac{\rho_{n}}{a^{n}}.
\end{align}
\subsubsection{$k \neq 0$}
\indent For a curved space results are found as
\begin{gather}
H(a)=\sqrt{\dfrac{8\pi G}{3}(\dfrac{\rho_{w}}{a}+\dfrac{\rho_{n}}{a^{n}})-\dfrac{k}{L^{2}a^{2}}},\\
V(a)=\dfrac{5\rho_{w}}{6a}+\dfrac{(6-n)}{6}\dfrac{\rho_{n}}{a^{n}},\\
\phi (a)=\pm\int_{a_{in}}^{a}\sqrt{\dfrac{n\rho_{n}a^{'}+\rho_{w}a^{'n}}{8\pi G[\rho_{w}a^{'2+n}+\rho_{n}a^{'3}]-3ka^{'1+n}}}da^{'}+\phi(a_{in}),\\
p(a)=\dfrac{-2\rho_{w}}{3a}+\dfrac{(n-3)}{3}\dfrac{\rho_{n}}{a^{n}},\\
q(a)=\dfrac{4\pi G[-\rho_{w}a^{n}+(n-2)\rho_{n}a]a}{8\pi G(\rho_{w}a^{n}+\rho_{n}a)a-3ka^{n}}.
\end{gather}
\\Behaviour of the deceleration parameter at the beginning is obtained as
\begin{align}
\lim _{a\rightarrow 0}q(a)=
\begin{cases}
-1+\dfrac{n}{2} \hspace{20pt}\text{if} \hspace{20pt} 2< n\leq 6,\\
0\hspace{20pt}\text{if} \hspace{20pt} 0< n\leq 2.
\end{cases}
\end{align}
\\Furthermore since denominator of $q(a)\sim H^{2}(a)$ as stated before, numerator of $q(a)$ determines dynamics of the universe. Roots of the deceleration parameter is found as
\begin{align}
a=[(n-2)\dfrac{\rho_{n}}{\rho_{w}}]^{1/(n-1)}.
\end{align}
\\Therefore the universe start to expand with deceleration and then turns to accelerate for $2<n\leq 6$. On the other hand for $0\leq n\leq 2$ at the beginning of universe $H(a)$ was constant and thus the universe starts to its expansion with acceleration.\\
\subsubsection{$k=0$}
\indent Behaviour of $q(a)$ changes as
\begin{align}
\lim _{a\rightarrow 0}q(a)=
\begin{cases}
-1+\dfrac{n}{2} \hspace{20pt}\text{if} \hspace{20pt} 1< n\leq 6,\\
-\dfrac{1}{2}\hspace{20pt}\text{if} \hspace{20pt} 0\leq n\leq 1.
\end{cases}
\end{align}
\\Therefore the differences in dynamics of the universe when it is spatially flat is seen when $0\leq n\leq 2$. In this case at the beginning $H(a)$ is not constant and the universe starts its expansion with acceleration.\\
\indent $\phi (a)$ has been simplified for these cases: domain walls and stiff fluid, domain walls and radiation, domain walls and matter as
\begin{align}
\phi(a)&=\pm\dfrac{1}{5\sqrt{2\pi G}}\{-\sqrt{6}ln[\dfrac{(\sqrt{6}\sqrt{\rho_{s}+\rho_{w}a^{5}}+\sqrt{6\rho_{s}+\rho_{w}a^{5}})a_{in}^{5/2}}{(\sqrt{6}\sqrt{\rho_{s}+\rho_{w}a_{in}^{5}}+\sqrt{6\rho_{s}+\rho_{w}a_{in}^{5}})a^{5/2}}]\}+ln[\dfrac{\sqrt{\rho_{s}+\rho_{w}a^{5}}+\sqrt{6\rho_{s}+\rho_{w}a^{5}}}{\sqrt{\rho_{s}+\rho_{w}a_{in}^{5}}+\sqrt{6\rho_{s}+\rho_{w}a_{in}^{5}}}]\nonumber \\
&+\phi(a_{in}) \hspace{20 pt}\text{if}\hspace{20pt}n=6, 
\end{align}
\begin{align}
\phi(a)&=\pm\dfrac{1}{3\sqrt{2\pi G}}\{-2ln[\dfrac{(2\sqrt{\rho_{r}+\rho_{w}a^{3}}+\sqrt{4\rho_{r}+\rho_{w}a^{3}})a_{in}^{3/2}}{(2\sqrt{\rho_{r}+\rho_{w}a_{in}^{3}}+\sqrt{4\rho_{r}+\rho_{w}a_{in}^{3}})a^{3/2}}]+ln[\dfrac{\sqrt{\rho_{r}+\rho_{w}a^{3}}+\sqrt{4\rho_{r}+\rho_{w}a^{3}}}{\sqrt{\rho_{r}+\rho_{w}a_{in}^{3}}+\sqrt{4\rho_{r}+\rho_{w}a_{in}^{3}}}]\}\nonumber \\
&+\phi(a_{in})\hspace{20 pt}\text{if}\hspace{20pt}n=4,
\end{align}
\begin{align}
\phi(a)&=\pm\dfrac{1}{2\sqrt{2\pi G}}\{-\sqrt{3}ln[\dfrac{(\sqrt{3}\sqrt{\rho_{m}+\rho_{w}a^{2}}+\sqrt{3\rho_{m}+\rho_{w}a^{2}})a_{in}}{(\sqrt{3}\sqrt{\rho_{m}+\rho_{w}a_{in}^{2}}+\sqrt{3\rho_{m}+\rho_{w}a_{in}^{2}})a}]+ln[\dfrac{\sqrt{\rho_{m}+\rho_{w}a^{2}}+\sqrt{3\rho_{m}+\rho_{w}a^{2}}}{\sqrt{\rho_{m}+\rho_{w}a_{in}^{2}}+\sqrt{3\rho_{m}+\rho_{w}a_{in}^{2}}}]\}\nonumber \\
&+\phi(a_{in})\hspace{20 pt}\text{if}\hspace{20pt}n=3.
\end{align}
\\It is possible to simply the relation between time and the scale factor as
\begin{gather}
t=\int_{a_{in}}^{a}\dfrac{da^{'}}{a^{'}H(a^{'})}, \nonumber \\
t=\int_{a_{in}}^{a}\dfrac{da^{'}}{a^{'}\sqrt{\dfrac{8\pi G}{3}(\dfrac{\rho_{w}}{a^{'}}+\dfrac{\rho_{n}}{a^{'n}})}},\\
t=\sqrt{\dfrac{3}{2\pi G\rho_{w}}}\Big[\sqrt{a^{'}}{}_{2}F_{1}(\dfrac{1}{2},\dfrac{1}{2-2n};\dfrac{1}{2-2n}+1;-\dfrac{\rho_{n}}{\rho_{w}}a^{'(1-n)})\Big]_{a_{in}}^{a}
\end{gather}
\\where ${}_{2}F_{1}$ is the hypergeometric function which was introduced in section 4.2.2. To avoid singularities in the hypergeometric function in (164) n should be chosen such that $3/2<n$.
\subsection{Dark Energy revisited}
\indent We have already examined the case $V=V_{0}$ in section 4.2. For spatially flat universe we have obtained 
\begin{gather}
a(t)=[\dfrac{b^{2}V_{0}e^{2\mu t}-\gamma(a_{in})a_{in}^{6}e^{-2\mu t}}{2bV_{0}}]^{1/3}, \nonumber \\
b=a_{in}^{3}+\sqrt{a_{in}^{6}+\dfrac{\gamma(a_{in})a_{in}^{6}}{V_{0}}} \hspace{20pt} \text{and} \hspace{20pt} \mu=\sqrt{6\pi GV_{0}} \nonumber.
\end{gather}
\\Then we formulate the Hubble function and the deceleration parameter as a function of  $t$ as
\begin{gather}
H(t)=\dfrac{2\mu}{3}(1-\dfrac{2f}{f-b^{2}V_{0}e^{4\mu t}}), \hspace{20pt} f=\gamma(a_{in})a_{in}^{6},\\
q(t)=-1+\dfrac{12b^{2}fV_{0}e^{4\mu t}}{(1+b^{2}V_{0}e^{4\mu t})^{2}}.
\end{gather}
\\First constraint on our parameters comes from positivity of the Hubble function
\begin{align}
H(t)>0 \hspace{20pt}\Rightarrow \hspace{20pt} f<b^{2}V_{0}.
\end{align}
\\On the other hand $q(t)$ has only one real root
\begin{align}
q(t)=0 \hspace{20pt} \Rightarrow \hspace{20pt} t=\dfrac{1}{4\mu}ln[\dfrac{(5+2\sqrt{6})f}{b^{2}V_{0}}].
\end{align}
\\We choose 
\begin{align}
t>0 \hspace{20pt} \Rightarrow \hspace{20pt} f>\dfrac{b^{2}V_{0}}{\tau}, \hspace{20pt} \tau=(5+2\sqrt{6}).
\end{align}
\\One can write
\begin{align}
f=(\dfrac{1}{\tau}+\varepsilon)b^{2}V_{0}, \hspace{15pt} 0<\varepsilon<1-\dfrac{1}{\tau}\hspace{10pt}\Rightarrow \dfrac{b^{2}V_{0}}{\tau}<f<b^{2}V_{0}.
\end{align}
\\Now we will find the condition which results in acceleration at the beginning of the universe
\begin{align}
\lim_{t->0}q(t)=-1+\dfrac{12b^{2}fV_{0}}{(f+b^{2}V_{0})^{2}} <0.
\end{align}
\\Thus
\begin{align}
5-\dfrac{1}{\tau}-\sqrt{24}<\varepsilon<1-\dfrac{1}{\tau},\\
4.3\times 10^{-16}<\varepsilon<0.90.
\end{align}
\\With these initial conditions universe starts to expand with acceleration and then turns into deceleration.
\subsection{Combination containing exotic matter}
\indent We have already explored the case of exotic matter in section 4.2. Now we will study combination of this kind of matter and some ordinary matters in the early universe. We have taken the most general form of the potential as
\begin{align}
V(a)=\dfrac{V_{s}}{a^{6}}+\dfrac{V_{n}}{a^{n}}.
\end{align}
\indent Related cosmological functions have been found as
\begin{gather}
\gamma(a)=\dfrac{1}{a^{6}}[\dfrac{nV_{n}}{6-n}a^{6-n}+6V_{s}ln(\dfrac{a}{a_{in}})+a_{in}^{6}\tilde{\gamma}(a_{in})], \hspace{15pt} \tilde{\gamma}(a_{in})=\gamma(a_{in})-\dfrac{nV_{n}}{(6-n)a_{in}^{n}}\\
H(a)=\sqrt{\dfrac{8\pi G}{3}[\dfrac{6V_{n}}{(6-n)a^{n}}+\dfrac{a_{in}^{6}\tilde{\gamma}(a_{in})+V_{s}[1+6ln(\dfrac{a}{a_{in}})]}{a^{6}}]-\dfrac{k}{a^{2}}},\\
\phi(a)=\pm\sqrt{6}\int_{a_{in}}^{a}\sqrt{\dfrac{(6-n)[a_{in}^{6}\tilde{\gamma}(a_{in})+6V_{s}ln(\dfrac{a^{'}}{a_{in}})]+nV_{n}a^{'6}}{(6-n)\{8\pi G[a_{in}^{6}\tilde{\gamma}(a_{in})+V_{s}(1+6ln(\dfrac{a^{'}}{a_{in}}))]-3ka^{'4}\}a^{'2+n}+48\pi GV_{n}a^{'8}}}da^{'}+\phi(a_{in}),
\end{gather}
\begin{gather}
\rho(a)=\dfrac{6V_{n}}{(6-n)a^{n}}+\dfrac{V_{s}}{a^{6}}ln[e(\dfrac{a}{a_{in}})^{6}]+\dfrac{a_{in}^{6}\tilde{\gamma}(a_{in})}{a^{6}},\\
p(a)=\dfrac{2(n-3)V_{n}}{(6-n)a^{n}}+\dfrac{V_{s}}{a^{6}}ln[\dfrac{1}{e}(\dfrac{a}{a_{in}})^{6}]+\dfrac{a_{in}^{6}\tilde{\gamma}(a_{in})}{a^{6}},\\
q(a)=\dfrac{8\pi G\{(6-n)[2a_{in}^{6}\tilde{\gamma}(a_{in})+V_{s}(-1+12ln(\dfrac{a}{a_{in}}))]a^{n}+3(n-2)V_{n}a^{6}\}}{(6-n)\{8\pi G[a_{in}^{6}\tilde{\gamma}(a_{in})+V_{s}(1+6ln(\dfrac{a}{a_{in}}))]-3ka^{4}\}a^{n}+48\pi GV_{n}a^{6}}.
\end{gather}
\subsubsection{Combination with radiation}
\indent When the exotic matter is accompanied with radiation its dynamics is governed by the following deceleration parameter
\begin{align}
q(a)=\dfrac{2a_{in}^{6}\tilde{\gamma}(a_{in})+V_{s}[-1+12ln(\dfrac{a}{a_{in}})]+3V_{r}a^{2}}{a_{in}^{6}\tilde{\gamma}(a_{in})+V_{s}[1+6ln(\dfrac{a}{a_{in}})]+3V_{r}a^{2}}, \hspace{15pt} \tilde{\gamma}(a_{in})=\gamma (a_{in})-\dfrac{2V_{r}}{a_{in}^{4}}.
\end{align}
\indent One can trace its behaviour back into time as
\begin{align}
q(a_{in})=\dfrac{2a_{in}^{6}\gamma(a_{in})-V_{r}a_{in}^{2}-V_{s}}{a_{in}^{6}\gamma(a_{in})+V_{r}a_{in}^{2}+V_{s}}.
\end{align}
\\The following condition
\begin{align}
V_{s}+V_{r}a_{in}^{2}>2a_{in}^{6}\gamma(a_{in}).
\end{align}
\\causes accelerated beginning for the universe. Moreover this choice also results in negative total pressure in the beginning as
\begin{align}
p(a_{in})=\dfrac{\gamma(a_{in})a_{in}^{6}-V_{r}a_{in}^{2}-V_{s}}{a_{in}^{6}}<0,
\end{align}
\\while energy density remains positive. Thus negative pressure results in accelerated motion for a while. Then this behaviour changes as pressure becomes positive and the universe decelerates. Therefore this exotic matter and radiation with initial condition which satisfies (183) also has mathematical turn on and turn off mechanism for accelerated motion in the early universe.
\subsubsection{Combination with domain walls} 
\indent In spatially flat universe the deceleration parameter becomes
\begin{align}
q(a)=\dfrac{10a_{in}^{6}\tilde{\gamma}(a_{in})-3v_{w}a^{5}+5v_{s}[-1+12ln(\dfrac{a}{a_{in}})]}{5a_{in}^{6}\tilde{\gamma}(a_{in})+6v_{w}a^{5}+5v_{s}[1+6ln(\dfrac{a}{a_{in}})]}.
\end{align}
\\Although this universe may start to expand with acceleration or deceleration, after a while it will accelerate because the leading term $a^{5}$ in the numerator has a negative coefficient.
\subsubsection{Combination with dark energy} 
\indent In spatially flat universe the deceleration parameter becomes
\begin{align}
q(a)=\dfrac{2a_{in}^{6}\tilde{\gamma}(a_{in})-v_{0}a^{6}+v_{s}[-1+12ln(\dfrac{a}{a_{in}})]}{a_{in}^{6}\tilde{\gamma}(a_{in})+v_{0}a^{6}+v_{s}[1+6ln(\dfrac{a}{a_{in}})]}.
\end{align}
\\Although this universe may start to expand with acceleration or deceleration, after a while it will accelerate because the leading term $a^{6}$ in the numerator has a negative coefficient.
\section{Late time expansion of the universe}
\indent We try to understand the present era which is usually described by dark matter and dark energy. However we consider different forms of energy components. Thus in the first subsection we model a universe with domain walls and matter. After obtaining formulas for functions we compare our results with supernova type Ia data just by curve fitting. In the second subsection we study dark energy dominated universe with the same steps. 
\subsection{Cosmic walls}
\indent The hypothesis that the scalar field is the dark matter and the dark energy was investigated for flat universe in \cite{matos2000scalar}. The results were compared with observations of type Ia supernovae which were available in 2000. In that study, matter part of the universe was neglected and it was found that $\rho_{\phi}\sim a^{-1.09}$ and $q_{0}=-0.45$. Different from them we include matter component of the universe and we solve field equations analytically. Then we compare our results with the type Ia supernovae data released in 2018 \cite{scolnic2018complete}. Furthermore in this comparision we extract the value of Hubble constant $H_{0}$ and the value of absolute magnitude $M$ with cosmological density parameters.\\
\indent Today our universe is believed to be almost flat and contribution of radiation to the energy density is very tiny. For this reason we investigate the case in which  
\begin{align}
\rho (a)=\dfrac{\rho _{m}}{a^{3}}+\dfrac{\rho _{w}}{a}.
\end{align}
\\Domain wall dominated universes have been already studied in section 5.2. Just plugging $n=3$ in section 5.2.2 we obtain the Hubble function, the scalar field and the potential as
\begin{align}
H(a)=\sqrt{\dfrac{8\pi G}{3}(\dfrac{\rho_{w}}{a}+\dfrac{\rho_{m}}{a^{3}})},
\end{align}
\begin{align}
\phi(a)&=\pm\dfrac{1}{2\sqrt{2\pi G}}\{-\sqrt{3}ln\Big [\dfrac{(\sqrt{3}\sqrt{\rho_{m}+\rho_{w}a^{2}}+\sqrt{3\rho_{m}+\rho_{w}a^{2}})a_{in}}{(\sqrt{3}\sqrt{\rho_{m}+\rho_{w}a_{in}^{2}}+\sqrt{3\rho_{m}+\rho_{w}a_{in}^{2}})a}\Big ]\nonumber \\&+ln\Big [\dfrac{\sqrt{\rho_{m}+\rho_{w}a^{2}}+\sqrt{3\rho_{m}+\rho_{w}a^{2}}}{\sqrt{\rho_{m}+\rho_{w}a_{in}^{2}}+\sqrt{3\rho_{m}+\rho_{w}a_{in}^{2}}}\Big ]\}+\phi(a_{in}),
\end{align}
\begin{gather}
V(a)=\dfrac{5\rho_{w}}{6a}+\dfrac{\rho_{m}}{2a^{3}},\\
p(a)=-\dfrac{2\rho _{w}}{3a},\\
q(a)=-\dfrac{1}{2}+\dfrac{\rho _{m}}{\rho _{m}+\rho _{w}a^{2}}, \\
t=\sqrt{\dfrac{3}{2\pi G\rho_{w}}}\Big[\sqrt{a^{'}}{}_{2}F_{1}(\dfrac{1}{2},-\dfrac{1}{4};\dfrac{3}{4};-\dfrac{\rho_{m}}{\rho_{w} a^{'2}})\Big]_{a_{in}}^{a}.
\end{gather}
\\Investigation of the deceleration parameter tells us
\begin{gather}
\lim_{a\rightarrow 0}q(a)=\dfrac{1}{2}, \hspace{20pt} \lim_{a\rightarrow \infty}q(a)=-\dfrac{1}{2} ,\\
q(a)=0 \hspace{20pt}\Rightarrow \hspace{20pt} a=\sqrt{\dfrac{\rho_{m}}{\rho_{w}}}.
\end{gather}
\\Hence if $\rho_{m}<\rho_{w}$ this model of the universe accelerates.\\ 
 \indent To test the reliability of the theoretical model we will use supernovae data. Luminosity distance-redshift relation had been already constructed as \cite{hobson2006general}
\begin{align}
d_{L}=(1+z)R_{0}S(\chi (z)).
\end{align} 
\\The comoving coordinate $\chi$ is given as
\begin{align}
\chi (z)=\dfrac{c}{R_{0}}\int _{0}^{z}\dfrac{dz^{'}}{H(z^{'})}.
\end{align}
\\The function $r=S(\chi)$ is given by
\begin{gather*}
S(\chi)=
\begin{cases}
Sin(\chi) & \text{if } k=1,  \nonumber \\
\chi & \text{if } k=0, \nonumber \\
Sinh(\chi)& \text{if } k=-1.
\end{cases}
\end{gather*}
\\Thus 
\begin{gather*}
R_{0}S(\chi (z)) =\dfrac{c}{H_{0}}   
\begin{cases}
 \left|\Omega _{k,0}\right |^{-1/2}S(\sqrt{\left|\Omega _{k,0}\right |}E(z)) & \text{for } \Omega_{k}\neq 0, \\     \nonumber \\
E(z) & \text{for } \Omega_{k}=0    
\end{cases}
\end{gather*}
\\ where $E(z)=\dfrac{R_{0}H_{0}}{c}\chi (z)$. 
\\For spatially flat universe with matter and domain walls it reduces to the following form
\begin{align}
d_{L}=\dfrac{c(1+z)}{H_{0}}\int _{0}^{z} \frac{dz^{'}}{[\Omega _{m}(1+z^{'})^{3}+\Omega _{w}(1+z^{'})]^{1/2}}
\end{align}
\\where $\Omega _{m}=1-\Omega _{w}$. The relation between observational measurements and the theory are established as  
\begin{align}
m=5log_{10}(\dfrac{d_{L}}{1Mpc})+25+M
\end{align}
\\where $m$ and $M$ are the apparent and the absolute magnitudes respectively. Then the distance modulus is defined as $\mu=m-M$. \\
\indent Before going further we would like to remind the Hubble tension and supernova absolute magnitude tension. Both of them are fundamental cosmological parameters. Their  values must be presented precisely.\\
\indent To determine the value of $H_{0}$ different methods have been applied. According to the Planck measurement of the cosmic microwave background  (CMB) anisotropies, assuming the base-$\Lambda$CDM model \cite{aghanim2020planck} 
$H_{0}=67.36\pm 0.54 kms^{-1}Mpc^{-1}$. On the other hand Hubble Space Telescope (HST) observations of Cepheids have been used to calibrate the measurements using type Ia supernovae in \cite{riess2019large} and it has been declared $H_{0}=74.13\pm 1.42 kms^{-1}Mpc^{-1}$.\\
\indent In last years it has been pointed out that the absolute peak magnitude $M_{B}$ of Type Ia supernovae is converted into a value of $H_{0}$ \cite{efstathiou2021h0,camarena2021use,nunes2021dark,colacco2022varying}. It's value has been stated as  $M_{B}=-19.401\pm 0.027$ mag \cite{camarena2020new} in 2020 and $M_{B}=-19.244\pm 0.037$ mag \cite{camarena2021use} in 2021 by application of different methods.\\
\indent The most recent data set for type Ia supernova observation wchich is called as Pantheon dataset was released in \cite{scolnic2018complete}. 1048 data points are presented as $(m,z)$ pairs where $z<2.3$. Since there is a debate on values of  $H_{0}$ and $M$ we include their values as parameters which are to be derived from curve fitting. Therefore we have to extract values of $\rho_{w}$, $H_{0}$ and $M$ from data. To be able to see effects of these numbers on each other separately we applied the curve fitting method for three combinations of two parameter sets. \\
\indent  Then we applied the $\chi ^{2}$ test to measure the goodness of these fits. $\chi ^{2}$ per degrees of freedom,  $\chi _{\nu}^{2}$ is calculated according to following formula
\begin{gather}
\chi ^{2}=\sum _{i}^{N}\dfrac{(\mu _{i}^{data}-\mu _{i}^{model})^{2}}{\sigma _{i}^{2}},\\ \nonumber\\
\chi ^{2}_{\nu}=\dfrac{\chi ^{2}}{\nu}, \hspace{20pt} \nu=N-k
\end{gather}
\\where $k$ is the number of parameters that will be extracted from the $N$ number of data points. \\ 
\indent First, we assign trial number for $\Omega_{w}$ and results are shown in Table I.
\begin{center}
    \begin{tabular}{ | l | l | l | l |}
    \hline
    $\Omega _{w}$ & $H_{0}$ & $M$ & $\chi^{2}/\nu$ \\ \hline
    0.70 & 72.0756 $\pm$ 0.0001 & -19.213 $\pm$ 0.004 & 1.124  \\ \hline
    0.75 & 72.0945 $\pm$ 0.0001 & -19.227 $\pm$ 0.004 & 1.075 \\ \hline
    0.80 & 72.1142 $\pm$ 0.0001 & -19.240 $\pm$ 0.004 & 1.038 \\ \hline
    0.85 & 72.1347 $\pm$ 0.0001 & -19.255 $\pm$ 0.004 & 1.014  \\ \hline
    0.90 & 72.1563 $\pm$ 0.0001 & -19.269 $\pm$ 0.004 & 1.009 \\ \hline
    0.95 & 72.1789 $\pm$ 0.0001 & -19.285 $\pm$ 0.004 & 1.026 \\ \hline
    0.9999 & 72.2028 $\pm$ 0.0001 & -19.301 $\pm$ 0.004 & 1.073 \\
    \hline
    \end{tabular}
\end{center}
TABLE I: Values of $H_{0}$ and $M$ for given $\Omega_{w}$\\ 
\indent The best fit which is obtained for  $\Omega _{w}=0.90$ gives $H_{0}=72.1563\pm 0.0001$, $M=-19.269\pm 0.004$ and $\chi^{2}/\nu=1.009$. \\
\indent After getting intuition about parameters we perform curve fitting for trial $M$ values. Results are given in Table II. \\
\begin{center}
    \begin{tabular}{ | l | l | l | l |}
    \hline
    $H_{0}$ & $M$ & $\Omega _{w}$ & $\chi^{2}/\nu$ \\ \hline
    74 & -19.211 $\pm$ 0.006 & 0.888 $\pm$ 0.015 & 1.008  \\ \hline
    73 & -19.240 $\pm$ 0.006 & 0.888 $\pm$ 0.015 & 1.008 \\ \hline
    72 & -19.270 $\pm$ 0.006 & 0.888 $\pm$ 0.015 & 1.008 \\ \hline
    71 & -19.301 $\pm$ 0.006 & 0.888 $\pm$ 0.015 & 1.008  \\ \hline
    70 & -19.332 $\pm$ 0.006 & 0.888 $\pm$ 0.015 & 1.008 \\ \hline
    69 & -19.363 $\pm$ 0.006 & 0.888 $\pm$ 0.015 & 1.008 \\ \hline
    68 & -19.395 $\pm$ 0.006 & 0.888 $\pm$ 0.015 & 1.008 \\ \hline
    67 & -19.427 $\pm$ 0.006 & 0.888 $\pm$ 0.015 & 1.008 \\
    \hline
    \end{tabular}
\end{center}
TABLE II: Values of $M$ and $\Omega_{w}$ for given $H_{0}$. \\ 
\indent Numbers in Table II tell us that the hundredths digit of $M$ is sensitively depended on the ones digit of $H_{0}$. In addition, more accurate result for $\Omega_{w}$ is obtained. All of the results have the same $\chi^{2}/\nu$ value. Thus we choose $3^{rd}$ line: $H=72$, $M=-19.270\pm 0.006$, $\Omega _{w}=0.888\pm 0.015$ and  $\chi^{2}/\nu=1.008$. These numbers are compatible with best fit of the table I.\\
\indent Effect of trial values of $M$ on $H_{0}$ and $\Omega_{w}$ are presented in Table III.\\
\begin{center}
    \begin{tabular}{ | l | l | l | l |}
    \hline
    $M$ & $H_{0}$ & $\Omega_{w}$ & $\chi^{2}/\nu$ \\ \hline
    -19.45 & 66.29 $\pm$ 0.19 & 0.889 $\pm$ 0.015 & 1.008  \\ \hline
    -19.40 & 67.83 $\pm$ 0.20 & 0.889 $\pm$ 0.015 & 1.008 \\ \hline
    -19.35 & 69.41 $\pm$ 0.20 & 0.889 $\pm$ 0.015 & 1.008 \\ \hline
    -19.30 & 71.03 $\pm$ 0.20 & 0.889 $\pm$ 0.015 & 1.008  \\ \hline
    -19.25 & 72.68 $\pm$ 0.21 & 0.889 $\pm$ 0.015 & 1.008 \\ \hline
    -19.20 & 74.38 $\pm$ 0.22 & 0.889 $\pm$ 0.015 & 1.008 \\ \hline
    -19.15 & 76.11 $\pm$ 0.22 & 0.889 $\pm$ 0.015 & 1.008 \\
    \hline
    \end{tabular} 
\end{center}
TABLE III: Values of $H_{0}$ and $\Omega_{w}$ for given $M$.\\ 
\indent It is apparent that the value of the ones digit of $H_{0}$ is sensitively depended on the hundredths digit of $M$. Our choice is the $5^{th}$ line: $M=19.25$, $H_{0}=72.68\pm 0.21$, $\Omega_{w}=0.889 \pm 0.015$ and $\chi^{2}/\nu=1.008$.  These numbers are in agreement with best fit of the table I.\\
\indent Assuming base $\Lambda$CDM cosmology, late universe parameters were found as $H_{0}=67.27\pm0.60$, $\Omega_{m}=0.3166\pm0.0084$ and $\Omega_{\Lambda}= 0.6834\pm0.0084$ in \cite{aghanim2020planck}. It is known that $\Omega_{m}=\Omega_{bm}+\Omega_{dm}$ where $\Omega_{bm}\simeq 0.05$ and $\Omega_{dm}\simeq 0.27$.  However our results indicate that late universe parameters as $H_{0}=72.68\pm0.21$, $\Omega_{w}=0.889\pm0.015$ and $\Omega_{m}=0.111\pm0.015$. Since $\Omega_{bm}\simeq 0.05$, $\Omega_{dm}\simeq 0.06$. Therefore domain wall dominated universe is a candidate to explain $94$ percentage of the structures in the present universe while still $6$ percentage of the universe remains as unknown. \\
\indent To compare our results with Pantheon-data graphically we draw distance modulus $\mu$ vs redshift $z$ plot. In Figure 1 we use results given in $5^{th}$ line of Table III where $M=-19.25$, $H_{0}=72.68\pm 0.21$, $\rho_{w}=0.889\pm 0.015$ and $\rho_{m}=0.111\pm 0.015$. \\
\begin{figure}[htb]
        \includegraphics[height=4in,width=6in]{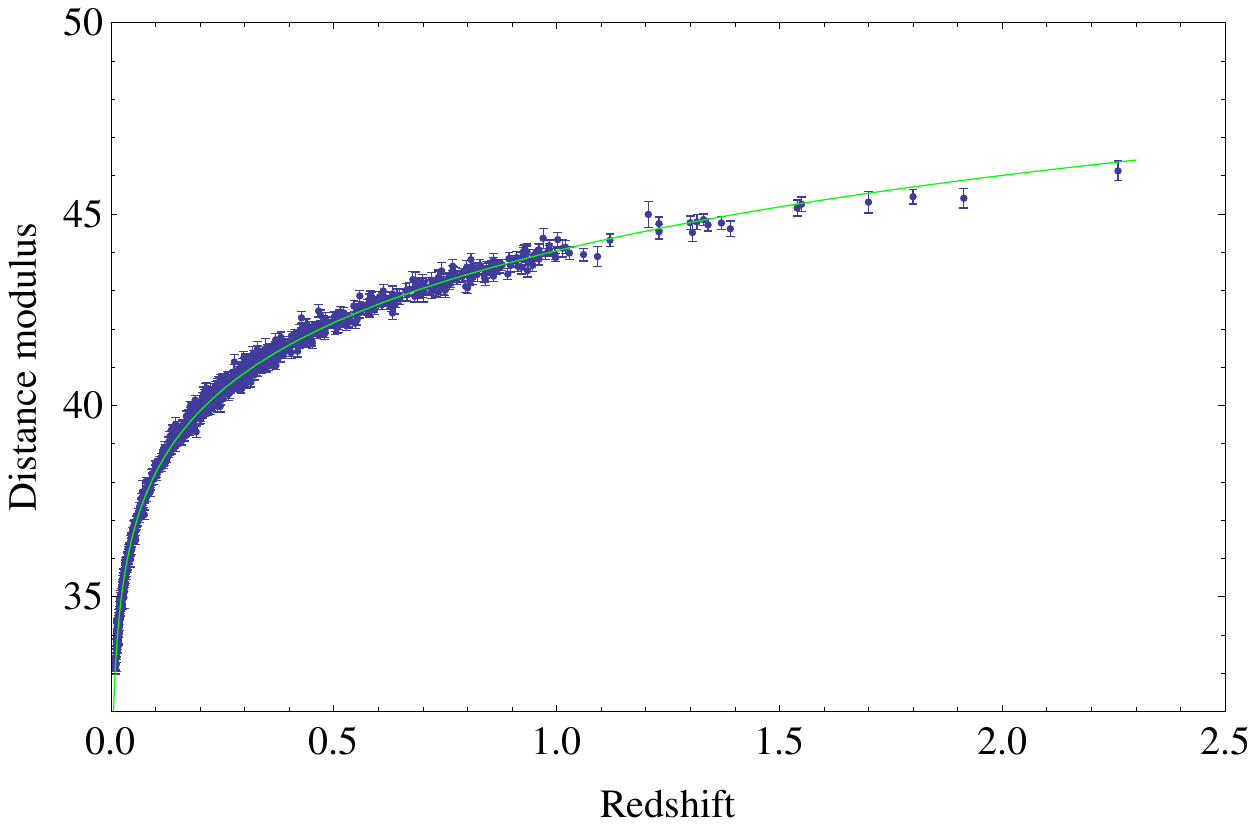}
     \caption{Figure I:Distance modulus vs redshift plot for domain wall dominated universe.\\Dots represent observation of Pantheon data while green line represents domain wall dominated universe }
  \end{figure}
\indent We obtain $q_{0}=-0.389$ by using values of cosmological density parameters in (192).
\subsection{Dark energy}
\indent Now we will present exact solutions for energy density given as
\begin{align}
\rho=\rho_{0}+\dfrac{\rho_{m}}{a^{3}}.
\end{align}
\\Actually this case corresponds to $n=3$ in section 5.1. For $k=0$ we have already obtained the scalar factor as
\begin{align}
a(t)=[\dfrac{(a_{in}^{3/2}+\sqrt{a_{in}^{3}+\dfrac{\rho_{m}}{\rho_{0}}})e^{3\mu t/2}+(a_{in}^{3/2}-\sqrt{a_{in}^{3}+\dfrac{\rho_{m}}{\rho_{0}}})e^{-3\mu t/2}}{2}]^{2/3}, 
\end{align}
\\where $\mu=\sqrt{\dfrac{8\pi G\rho_{0}}{3}}$. In addition potential is formulated as
\begin{align}
V(\psi)=\dfrac{1}{2}\dfrac{\rho_{m}}{a^{3}(\psi)}+\rho_{0},
\end{align}
\begin{gather}
a(\psi)=2^{2/3}\{\dfrac{a_{in}^{3/2}e^{\sqrt{3}\psi}[-1+\sqrt{1+a_{in}^{3}\dfrac{\rho_{0}}{\rho_{m}}}-(1+\sqrt{1+a_{in}^{3}\dfrac{\rho_{0}}{\rho_{m}}})e^{2\sqrt{3}\psi}]}{a_{in}^{3}\dfrac{\rho_{0}}{\rho_{m}}(1+e^{4\sqrt{3}\psi})-2(2+a_{in}^{3}\dfrac{\rho_{0}}{\rho_{m}})e^{2\sqrt{3}\psi}}\}^{2/3}, \\
\psi=\pm\sqrt{2\pi G}[\phi(a)-\phi(a_{in})].
\end{gather}
\indent Behaviour of the deceleration parameter is shown by
\begin{gather}
q(a)=\dfrac{-2\rho_{0}a^{3}+\rho_{m}}{2(\rho_{0}a^{3}+\rho_{m})}, \\
\hspace{20pt} \lim_{a\rightarrow \infty}q(a)=-1.
\end{gather}
\\To extract cosmological parameters from Pantheon data we apply the procedure as explained in the previous subsection with modification 
\begin{align}
d_{L}=\dfrac{c(1+z)}{H_{0}}\int _{0}^{z} \frac{dz^{'}}{[\Omega _{m}(1+z^{'})^{3}+\Omega _{0}]^{1/2}}
\end{align}
\\where $\Omega_{m}=1-\Omega_{0}$. \\
\indent First, we perform curve fitting for trial values of $\Omega_{0}$. Results are shown in Table IV.
\begin{center}
    \begin{tabular}{ | l | l | l | l |}
    \hline
    $\Omega _{0}$ & $H_{0}$ & $M$ & $\chi^{2}/\nu$ \\ \hline
    0.50 & 72.0674 $\pm$ 0.0001 & -19.208 $\pm$ 0.004 & 1.199  \\ \hline
    0.60 & 72.1183 $\pm$ 0.0001 & -19.243 $\pm$ 0.004 & 1.059 \\ \hline
    0.70 & 72.1749 $\pm$ 0.0001 & -19.282 $\pm$ 0.004 & 0.992 \\ \hline
    0.80 & 72.2384 $\pm$ 0.0001 & -19.324 $\pm$ 0.004 & 1.041  \\ \hline
    0.90 & 72.3116 $\pm$ 0.0001 & -19.372 $\pm$ 0.005 & 1.296 \\ 
    \hline
    \end{tabular}
\end{center}
TABLE IV: Values of $H_{0}$ and $M$ for given $\Omega_{0}$\\ \\
\indent The best fit occurs for $\Omega_{0}=0.70$. Thus $H_{0}=72.1749\pm 0.0001$, $M=-19.282\pm 0.004$ and $\chi^{2}/\nu=0.992$. \\
\indent Then we perform curve fitting for trial $M$ values. Results are given in Table V. \\
\begin{center}
    \begin{tabular}{ | l | l | l | l |}
    \hline
    $H_{0}$ & $M$ & $\Omega _{0}$ & $\chi^{2}/\nu$ \\ \hline
    74 & -19.234 $\pm$ 0.007 & 0.715 $\pm$ 0.012 & 0.990  \\ \hline
    73 & -19.264 $\pm$ 0.007 & 0.715 $\pm$ 0.012 & 0.990 \\ \hline
    72 & -19.294 $\pm$ 0.007 & 0.715 $\pm$ 0.012 & 0.990 \\ \hline
    71 & -19.324 $\pm$ 0.007 & 0.715 $\pm$ 0.012 & 0.990  \\ \hline
    70 & -19.355 $\pm$ 0.007 & 0.715 $\pm$ 0.012 & 0.990 \\ \hline
    69 & -19.386 $\pm$ 0.007 & 0.715 $\pm$ 0.012 & 0.990 \\ \hline
    68 & -19.418 $\pm$ 0.007 & 0.715 $\pm$ 0.012 & 0.990 \\ \hline
    67 & -19.450 $\pm$ 0.007 & 0.715 $\pm$ 0.012 & 0.990 \\
    \hline
    \end{tabular}
\end{center}
TABLE V: Values of $M$ and $\Omega_{0}$ for given $H_{0}$. \\ \\
\indent Results in the third line are compatible with the best fit of Table IV. $M= -19.294 \pm 0.007$, $\Omega_{0}=0.715\pm 0.012$ and $\chi^{2}/\nu=0.990$ are obtained for a given $H_{0}=72$.\\
\indent Finally we test the effect of $M$ on $H_{0}$ and $\Omega_{0}$. Results are presented in Table VI.\\
\begin{center}
    \begin{tabular}{ | l | l | l | l |}
    \hline
    $M$ & $H_{0}$ & $\Omega_{0}$ & $\chi^{2}/\nu$ \\ \hline
    -19.45 & 67.00 $\pm$ 0.21 & 0.715 $\pm$ 0.012 & 0.990   \\ \hline
    -19.40 & 68.57 $\pm$ 0.21 & 0.715 $\pm$ 0.012 & 0.990  \\ \hline
    -19.35 & 70.16 $\pm$ 0.22 & 0.715 $\pm$ 0.012 & 0.990  \\ \hline
    -19.30 & 71.80 $\pm$ 0.22 & 0.715 $\pm$ 0.012 & 0.990   \\ \hline
    -19.25 & 73.47 $\pm$ 0.23 & 0.715 $\pm$ 0.012 & 0.990  \\ \hline
    -19.20 & 75.18 $\pm$ 0.23 & 0.715 $\pm$ 0.012 & 0.990  \\ \hline
    -19.15 & 76.93 $\pm$ 0.24 & 0.715 $\pm$ 0.012 & 0.990  \\
    \hline
    \end{tabular} 
\end{center}
TABLE VI: Values of $H_{0}$ and $\Omega_{w}$ for given $M$.\\ 
\indent Numbers in the fourth line are compatible with the best fit of Table IV. $H_{0}=71.80\pm 0.22$, $\Omega_{0}=0.715\pm 0.012$ and $\chi^{2}/\nu=0.990$ are found for a given $M=-19.30$.\\
\indent All the tables in this article have a common interpretation: The number in ones digit of the Hubble constant $H_{0}$ is sensitively depended on the number in the hundredths digit of the absolute magnitude $M$ in both models. In addition as the value of $H_{0}$ increases, the value of $M$ decreases.  We will stop to dig more about this argument here since it is beyond the scope of our article. Cosmologists will continue to reveal the relation between $H_{0}$ and $M$ more clearly on further studies.\\
\indent Now we would like to compare our results for a domain wall dominated universe and a dark energy dominated universe in the same plot. However this goal can not be achieved accurately, because best fit values of $M$ are different for both models. For this reason we plot two figures.\\
\indent We draw our Figure 2 by taking $M=-19.25$ for which one of the best fits of the domain wall dominated universe is obtained with parameters $H_{0}=72.68\pm 0.21$, $\Omega_{w}=0.889\pm 0.015$ and $\Omega_{m}=0.111\pm 0.015$. On the other hand we have obtained $H_{0}=73.47\pm0.23$, $\Omega_{0}=0.715\pm0.012$, $\Omega_{m}=0.285\pm0.12$ and $\chi^{2}/\nu=0.990$  for $M=-19.25$.\\
\indent In Figure 3 we choose one of the best fits of the dark energy dominated universe for which $M=-19.30$, $H_{0}=71.80\pm0.22$, $\Omega_{0}=0.715\pm0.012$ and $\Omega_{m}=0.285\pm0.012$. In the first model we have obtained $H_{0}=71.03\pm0.20$, $\Omega_{w}=0.889\pm0.015$, $\Omega_{m}=0.111\pm0.015$ and $\chi^{2}/\nu=0.990=1.008$ for $M=-19.30$\\
\begin{figure}[htb]
   \begin{minipage}{0.58\textwidth}
     \includegraphics[height=2in,width=3in]{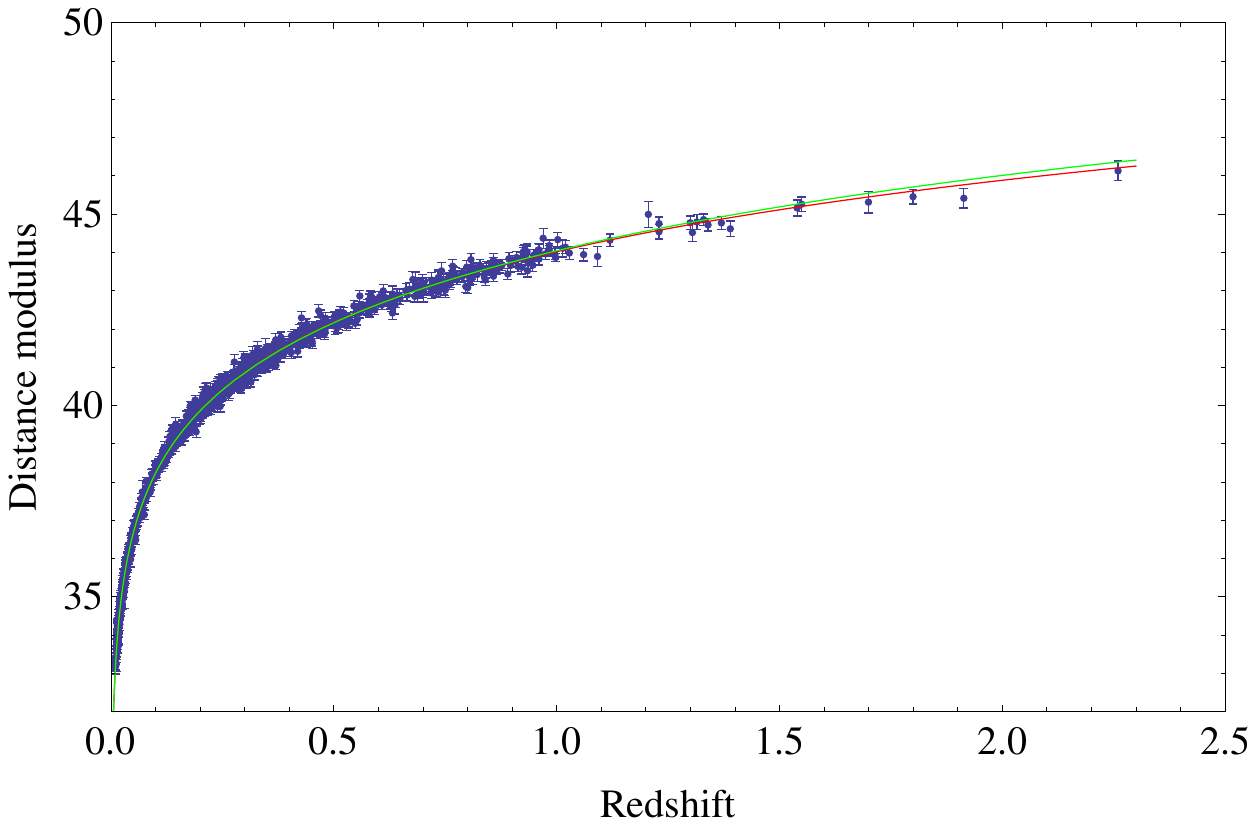}
     \caption{Distance modulus vs redshift plot\\ for $M=-19.25$. Dots represent observations,\\ green line represents domain wall dominated \\universe and red line represents dark energy \\ dominated universe.}
   \end{minipage}
   \begin {minipage}{0.58\textwidth}
     \includegraphics[height=2in,width=3in]{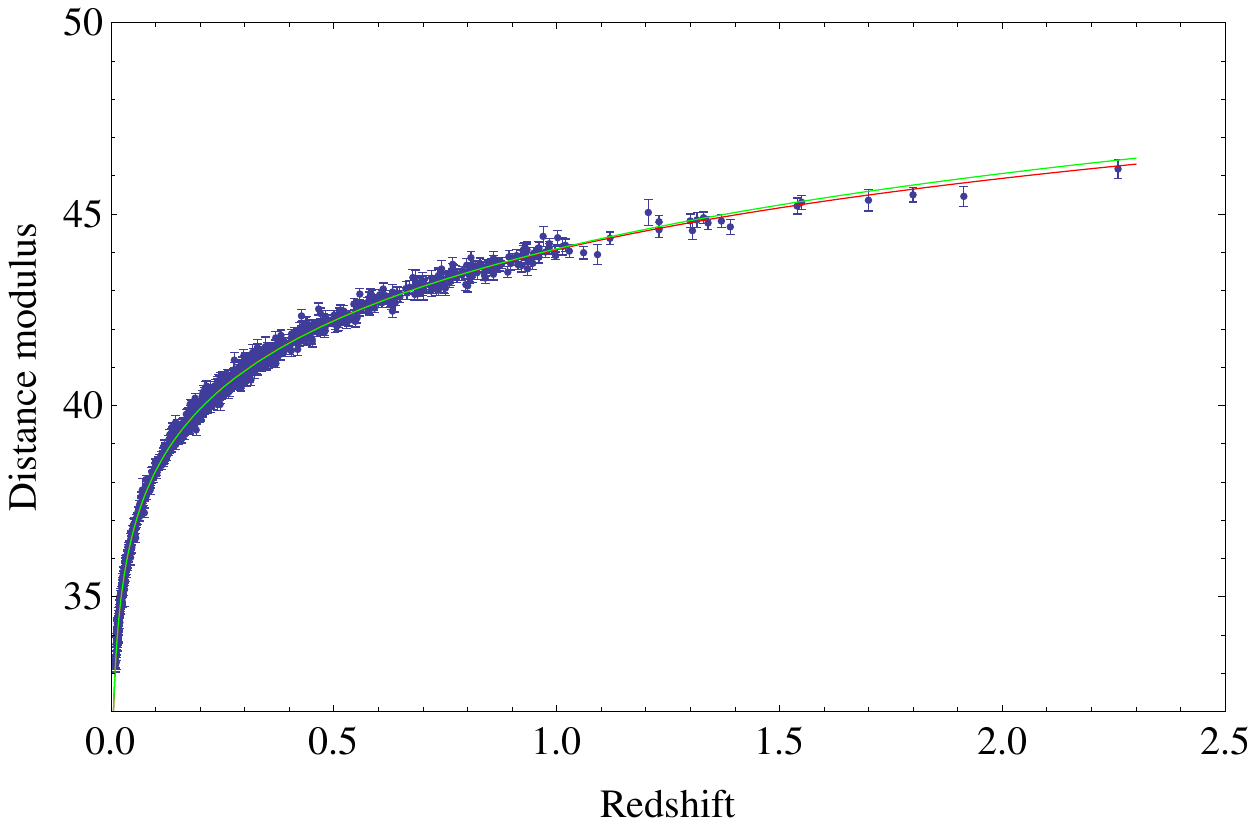}
     \caption{Distance modulus vs redshift plot\\ for $M=-19.30$. Dots represent observations,\\ green line represents domain wall dominated \\universe and red line represents dark energy \\ dominated universe.}
   \end{minipage}
\end{figure}
\indent Two figures are almost the same because $\chi^{2}/\nu$ for both models are very close the $1$. We need more data for bigger redshift values to decide whether one of the models is superior to the other one. We obtain $q_{0}=0.572$ by using these values of cosmological parameters in (207).
\section{Conclusion}
\indent We studied FLRW cosmology with real scalar field which is minimally coupled to gravity. Our main motivation in this article has been to assign the effective scalar field a source of all components of energy density. We applied a change of variable twice which is a more powerful method in the set of differential equations which represent dynamics of the universe. In the first one, we have replaced the independent variable $"t"$ with $"a"$. In the second one, we have changed the dependent variable of the $\phi$  equation so that it became a first order linear differential equation. We presented exact solutions in four different forms; solutions for given $V(a)$, solutions for given $\phi(a)^{'}$, solutions for given $H(a)$, and solutions for given $\rho(a)$. \\
\indent Then we have examined  single component universes and two component universes for a given energy density and for a given potential. In these solutions we have searched for mathematical mechanisms which create turn on and turn off for early inflationary expansion. We have explored mathematical structure of a new exotic matter. Equation of state for this component changes with the scale factor or equivalently changes with time. A universe which consists of radiation and this exotic matter, has mathematical machinery to turn on and to turn off inflationary expansion in early epoch.\\
\indent We have investigated the present era of the universe. Domain wall dominated universe and dark energy dominated universe have been studied. We have extracted numerical values of cosmological parameters from the most recent type Ia supernova data by taking care of the Hubble tension and the absolute magnitude tension. For domain wall dominated universe we have found that $\Omega_{w}=0.889\pm0.015$, $\Omega_{m}=0.111\pm0.015$ and $H_{0}=72.68\pm0.21$ for $M=-19.25$. This universe accelerates with $q_{0}=-0.389$. \\
\indent On the other hand for dark energy dominated universe cosmological parameters have been found as $\Omega_{0}=0.715\pm0.012$, $\Omega_{m}=0.285\pm0.012$ and $H_{0}=71.80\pm0.22$ for $M=-19.30$. Deceleration parameter of this universe is $q=-0.572$. Detailed analysis for the relation between distance modulus and redshift have shown that the number in ones digit of the Hubble constant $H_{0}$ is sensitively depended on the number in the hundredths digit of the absolute magnitude $M$ in both models. \\
\indent These two analyses indicate that both models equivalently explains dynamics of late time accelerated expansion of the universe. The difference between these models most probably will be seen when bigger redshift data are available.\\
\section*{Acknowledgement}
\indent We would like to acknowledge fruitful discussion about observations of type Ia supernovae with Aşkın Ankay, Önder Dünya and Kazım Çamlıbel. We thank Bogazici University for the financial support provided by the Scientific Research Fund (BAP), research Project No. 16521.

\appendix
\renewcommand{\theequation}{A-\arabic{equation}}      
  \setcounter{equation}{0}  

\section*{Appendix}
\subsection*{Change of variable}

\begin{align}
H(a)=\dfrac{\dot{a}}{a}
\end{align}
Therefore our new independent variable becomes a scale factor $'a'$.For this reason we write all other variables in terms of the new variable.
\begin{align}
\phi=\phi(a) \hspace{20pt} V(\phi)=V(a) \hspace{20pt} \dot{a}=aH(a)
\end{align}
\\As a result we obtain by change of variable
\begin{gather}
\dfrac{d\phi}{dt}=\dfrac{d\phi(a)}{dt}\dfrac{da}{dt}=\phi^{'}\dot{a}=\phi^{'}aH \\ \nonumber\\
\dot{\phi}=\phi^{'}aH \\ \nonumber \\
\dfrac{d^{2}\phi}{dt^{2}}=\dfrac{d}{dt}(\phi^{'}aH)=\dfrac{d}{da}(\phi^{'}aH)\dfrac{da}{dt}\\ \nonumber\\
\ddot{\phi}=\phi^{''}a^{2}H^{2}+\phi^{'}aH^{2}+\phi^{'}a^{2}HH^{'}
\end{gather}
\\By the help of the chain rule
\begin{align}
\dfrac{dV}{d\phi}=\dfrac{dV}{da}\dfrac{da}{d\phi}=\dfrac{dV}{da}\dfrac{1}{\dfrac{d\phi}{da}} , \\ \nonumber\\
\dfrac{dV}{d\phi}=V^{'}\dfrac{1}{\phi^{'}} .
\end{align}
\ On the other hand by starting from our definition we get followings
\begin{gather}
\dfrac{\dot{a}}{a}=H(a) \\ \nonumber\\
\dfrac{\ddot{a}}{a}-\dfrac{\dot{a}^{2}}{a^{2}}=\dfrac{dH}{da}\dfrac{da}{dt}\\\nonumber \\
\dfrac{\ddot{a}}{a}=H^{'}aH+H^{2}
\end{gather}
\subsection*{Integration by parts}
By using functions $u$ and $v$ a theorem integration by parts is written as
\begin{align}
\int_{x_{0}}^{x} u(x)dv(x)=u(x)v(x)\Biggr|_{x_{0}}^{x}-\int_{x_{0}}^{x} v(x)du(x)
\end{align} 
\\When calculating the function $\gamma (a)$ we choose
\begin{align}
u=-a^{'6} \hspace{20pt} dv=V^{'}da^{'}
\end{align}
\\thus
\begin{align}
du=-6a^{'5}da^{'}, \hspace{20pt} v=V(a^{'})
\end{align}
\\results in
\begin{align}
\int_{a_{in}}^{a} (-a^{'6}V^{'})da^{'}&=-a^{'6}V(a^{'})\Biggr|_{a_{in}}^{a}+6\int_{a_{in}}^{a} V(a^{'})a^{'5}da^{'} \nonumber \\
\int_{a_{in}}^{a} (-a^{'6}V^{'})da^{'}&=-a^{6}V(a)+6\int_{a_{in}}^{a} V(a^{'})a^{'5}da^{'}+a_{in}^{6}V(a_{in})
\end{align}

\bibliographystyle{unsrt}
\bibliography{bibfile}
\end{document}